\documentclass[10]{article}

\usepackage{subfig}
 \usepackage[margin=0.95in]{geometry}
 \usepackage{subcaption}

\usepackage{amsmath}
\usepackage{amssymb}
\usepackage{graphicx}
\usepackage{amsfonts}
\usepackage{tabularx}
\usepackage{booktabs}%
\providecommand{\U}[1]{\protect\rule{.1in}{.1in}}
\providecommand{\U}[1]{\protect\rule{.1in}{.1in}}

\title{ \bf From Agent-Based Markov Dynamics to Hierarchical Closures on Networks: Emergent Complexity and Epidemic Applications}

\author{ \it \small {\bf Entropy} 2026, 28(1), 63;  \\
\it \small https://doi.org/10.3390/e28010063 \\
\\ A. Y. Klimenko, A. Rozycki, Y. Lu \\  
\\ \it Centre For Multiscale Energy Systems, \\
\it School of Mechanical and Mining Engineering, \\  
\it The University of Queensland, St. Lucia 4072, Australia \\
email: a.klimenko@uq.edu.au }

\begin{document}


\maketitle

\begin{abstract}
We explore a rigorous formulation of agent-based SIR epidemic dynamics as a
discrete-state Markov process, capturing the stochastic propagation of
infection or an invading agent on networks. Using indicator functions and
corresponding marginal probabilities, we derive a hierarchy of evolution
equations that resembles the classical BBGKY hierarchy in statistical
mechanics. The structure of these equations clarifies the challenges of
closure and highlights the principal problem of systemic complexity arising
from stochastic but generally not fully chaotic interactions. Monte Carlo
simulations are used to validate simplified closures and approximations,
offering a unified perspective on the interplay between network topology,
stochasticity, and infection dynamics. We also explore the impact of lockdown
measures within a networked agent framework, illustrating how SIR dynamics and
structural complexity of the network shape epidemic with propagation of
COVID-19 in Northern Italy taken as an example.


\end{abstract}


\section{Introduction}

Understanding the stochastic dynamics  of epidemics, particularly those
involving competitive propagation, remains a central challenge not only in
epidemiology but also in related fields such as the spread of dominant species
or technological innovations. The topology of the invaded space can often be
effectively represented as a network, introducing additional complexity into
the dynamics of the epidemic. The classical SIR
(Susceptible/Infected/Recovered) model forms the basis for many such studies
and has been extended to incorporate realistic transmission patterns using
network-based formulations \cite{kiss2017mathematics}. Clear specification and
relative simplicity is a significant advantage of SIR as a standard model
capturing principal physical processes. Agent-based models (ABMs) offer a
powerful but computationally expensive framework for evaluating the overall
dynamics based on individual-level interactions
\cite{keeling2005networks,pastor2015epidemic}, --- these models conceptually
replicate particle approaches in modelling of reacting flows \cite{POPE1985}.

In this work, we adopt a probabilistic and systemic perspective by modelling
agent-based SIR dynamics as a continuous-time Markov process. Each individual
is treated as a node in a graph, whose state evolves due to infection and
recovery events governed by stochastic rules and controlled by the
corresponding master (Kolmogorov) equations. This framework allows for a
rigorous derivation of time-dependent joint and marginal probability
distributions that describe the transmission.

Our formulation follows the physics-based approach of Omata \cite{Omata2017},
but departs from traditional moment-based closures by deriving the governing
equations directly from the indicator functions associated with individual
node states. Taking ensemble averages of products of these indicators yields
exact evolution equations for marginal probabilities. The resulting structure
forms an explicit and interpretable hierarchy: equations for low-order
marginals depend on higher-order marginals because infection events couple the
stochastic states of neighbouring nodes.

This hierarchy is closely related in spirit to the
Bogoliubov--Born--Green--Kirkwood--Yvon (BBGKY) hierarchy in statistical
mechanics \cite{Yvon1935,BornGreen1946,Kirkwood1946,bogoliubov1946kinetic}. In
kinetic theory, the BBGKY hierarchy arises when the high-dimensional Liouville
equation is reduced to one- or few-particle marginal distributions:
interactions ensure that each reduced equation involves higher-order
distribution functions and is therefore unclosed at any fixed level.
Boltzmann's closure becomes possible only under additional assumptions, most
notably the molecular-chaos (Stosszahlansatz) hypothesis
\cite{Boltzmann1872,ChapmanCowling1970}. A closely related issue arises in
particle-based or agent-based models: low-order descriptions close easily only
under \textquotedblleft chaotic\textquotedblright\ assumptions, whereas
departures from chaos correspond to the emergence of non-trivial correlations
and, more broadly, to the emergence of complexity
\cite{klimenko2009lagrangian,KlimenkoPope2012CTM,klimenko2019evolution}.

In the present work, rather than introducing heuristic closures at the outset,
we retain the hierarchy in a symbolic and formally exact form, making the
dependence on higher-order stochastic structure explicit. In contrast to many
network-epidemic models based on expected values and deterministic ODEs, this
formulation preserves the probabilistic content of the underlying
continuous-time Markov jump process. Generalised derivatives provide a
convenient calculus for jump processes (infection and recovery events) and
connect naturally to ensemble-based Monte Carlo realisations
\cite{gillespie1977exact}, thereby unifying analytical derivations and
numerical simulations within a single framework.

Finally, real epidemics and realistic contact structures involve interventions
(e.g.\ lockdown-type reductions in transmission), behavioural adaptation, and
pronounced network heterogeneity. To probe these effects within the same
modelling framework, we examine how intervention timing and intensity interact
with clustered network structure to shape epidemic propagation. This
complements recent work on nonlinear outcomes in temporal and adaptive
networks \cite{holme2012temporal,block2020social} and highlights how
structural constraints and stochastic transmission jointly govern multi-wave
and long-tailed dynamics. Our main aim, however, is methodological: to model and 
analyse complex effects, rather than to deliver a comprehensive representation of realistic epidemic. 

Section 2 introduces the agent-based epidemic model as a continuous-time
Markov process and defines its probabilistic structure using indicator
functions and marginal distributions. Section 3 is dedicated to deriving the
governing equations for fine-grained and marginal probabilities, revealing a
BBGKY-like hierarchy. In Section 4, several closure strategies are proposed to
make this hierarchy tractable. Section 5 benchmarks these closures against
exact solutions and Monte Carlo simulations on simple graphs. Section 6
extends the analysis to randomly constructed networks, while Section 7 summarises the key findings.
The Appendix provides an example application of the model, together with a nomenclature list.

\section{Agent-based epidemic modelling as a Markov process.}

\subsection{System states and their full joint probability distribution}

In general, agent-based models involve two principal categories of agents:
nodes, which remain stationary, and particles, which can move from one node to
another. Both categories of agents can possess properties that may evolve in
time and/or change due to the interactions with other agents. In this work, we
focus on the interpretation of an epidemic model that represents individuals as nodes
numbered $i=1,2,$..., $N.$ Each node $i$ has a property $Y_{i}$ that can take
several values. According to the traditional SIR (susceptible, infected,
recovered) model, $Y_{i}$ can take one of the values S, I, or R. Therefore the
state of the system of nodes is given by the following vector
\begin{equation}
\mathbf{Y}^{(N)}=\left[  Y_{1},Y_{2},...,Y_{N}\right]  \ .
\end{equation}
For example S$_{1}$R$_{2}$R$_{3}$I$_{4}$,...,I$_{N-1}$,S$_{N}$ is a possible
state of the system, where nodes 1 and $N$ are susceptible, nodes 4 and $N-1$
are infected, and nodes 2 and 3 have recovered. There are 3$^{N}$ possible
states for this system. While more sophisticated models, which, for example,
may involve several infected states I$^{(1)},$I$^{(2)},...$ can be formulated
for specific diseases and our analysis can be easily extended to such models,
we prefer to keep our consideration general and focus on complexity emerging
at systemic levels. We take a systemic perspective and are interested in
general conceptual properties rather than in a detailed description of a
specific infection.

The propagation of an epidemic is, evidently, a random process which can be
characterised by the corresponding \textit{joint probabilities} $P_{\mathbf{Y}%
}^{(N)}=P(Y_{1}^{\circ},Y_{2}^{\circ},...,Y_{N}^{\circ})$ that can be
expressed as the following ensemble average
\begin{equation}
P_{\mathbf{Y}}=P^{(N)}=P\left(  Y_{1}^{\circ},Y_{2}^{\circ},...,Y_{N}^{\circ
}\right)  =\left\langle \theta_{1}(Y_{1}^{\circ})\theta_{2}(Y_{2}^{\circ
})...\theta_{N}(Y_{N}^{\circ})\right\rangle
\end{equation}
of the indicator functions
\begin{equation}
\theta_{i}(Y^{\circ})=\delta_{Y_{i}Y^{\circ}}=\left\{
\begin{array}
[c]{cc}%
1, & Y_{i}=Y^{\circ}\\
0, & Y_{i}\neq Y^{\circ}%
\end{array}
\right.  \ . \label{IndFun}%
\end{equation}
Here, \ $\delta$ denotes the Kronecker delta, while $\theta_{i}(Y^{\circ})$ is
a stochastic function that depends on location $i$ and the sample-space
parameter $Y^{\circ},$ which can take one of the three values $\{$S$,$%
I$,$R$\}.$ We use the superscript "$\circ$" to distinguish a random value
$Y_{i}(t),$ which is the actual state of the node $i$ at a given time moment
$t,$ from the corresponding sample-space parameter $Y_{i}^{\circ},$ which does
not depend on time and is an argument of the function $\theta_{i}(...)$. For
example if $Y_{i}=$I$,$ then $\theta_{i}($I$)=1$ and $\theta_{i}($%
S$)=\theta_{i}($R$)=0$. Note that the indicator functions depend on time
$\theta_{i}(Y_{i}^{\circ})=\theta_{i}(Y_{i}^{\circ},t)$ since $Y_{i}=Y_{i}(t)$
in definition (\ref{IndFun}). The complete probability function $P_{\mathbf{Y}%
}$ depends on $N$ sample space parameters $Y_{1}^{\circ},Y_{2}^{\circ
},...,Y_{N}^{\circ}$, and each of these parameters can independently take of
the three values: S, I or R. Note that the order of the nodes $1,...,i,...,N$
is deemed to be fixed to avoid confusion.

\subsection{Agent-based models as networks}

The SIR model involves interactions between individuals that propagate
infection from one individual to another. Similar mechanisms are engaged in
the transmission of ideas, news or other types of information between
individuals. These individuals are represented by nodes, which in addition to
properties $Y_{i},$ are characterised by connections to other nodes indicating
possible routes for transmission of the infection (or information). Note that
infection can propagate in both directions (i.e. from \ $i$ to $j$ and from
$j$ to $i$ but only if nodes $i$ and $j$ are connected). Hence, from the
mathematical perspective, the system of nodes is an undirected graph or
network. The adjacency matrix associated with this graph is denoted by
$A_{ij}$ --- this matrix has positive values if and only if
\ $i\leftrightarrow j$ (i.e. if nodes $i$ and $j$ are connected) . Note that
the adjacency matrix is symmetric $A_{ij}=A_{ji}$ for undirected graphs and,
conventionally, $A_{ii}=0.$ Representing interactions between individuals by
graphs is effective since each individual usually has relatively few direct
contacts, while contacts with the rest of the population are absent or
negligible. When using graphs, we avoid consideration of interactions between
nodes that do not interact. Graphs and networks are characterised by the
overall number of nodes $N$ and the overall number of edges $E.$ In real-world
networks, the number of nodes $N$ can reach millions, while the number of
edges is much smaller than its maximal value $N\ll N_{\max}=n(n-1)/2$. Two
classes of graphs can be considered: weighted and unweighted. For unweighted
graphs, the nodes $i$ and $j$ are either connected $A_{ij}=1$ or not
$A_{ij}=0$. In weighted graphs, each positive values $A_{ij}$ reflect the
intensity of connections between nodes $i$ and $j:$
\begin{equation}
A_{ij}=A_{ji}=\left\{
\begin{array}
[c]{cc}%
>0 & i\leftrightarrow j\text{ }\\
=0 & i\nleftrightarrow j
\end{array}
\right.  ,\ \ \ \ i,j=1,2,...,N\ .
\end{equation}

While it is natural to use networks to represent contacts and communications
between individuals, the properties of these networks evolved in modern
society. While networks of the past were subject to localisation determined by
physical distances, modern technology largely removed these constraints
allowing for effective communications and fast transportation. These modern
networks have so-called small-world properties: the number of nodes $N_{r}$
located within distance $r$ (measured in the minimal number of edges required
to pass while moving from one node to another) increases exponentially with
$r$
\begin{equation}
N_{r}\sim\exp(r),
\end{equation}
which is much faster than, say, the estimate\ $N_{r}\sim r^{2}$ that is valid
for a network localised on a two-dimensional surface.\ The modern world is
highly interconnected, creating favourable conditions not only for the
exchange of knowledge and information but also for the spread of infections.
Such spread remains, to a large extent, diffusive in character, being driven
by a multitude of local contacts \cite{Kendall1956}. However, occasional
long-distance \textquotedblleft jumps\textquotedblright\ can substantially
accelerate transmission, as illustrated by the small-world network phenomenon.

\subsection{The forward Kolmogorov equation}

From the perspective of the probability theory, the evolution of the system of
nodes is a Markov chain. In simple terms, the Markov property implies that
given the complete present state, we do not need to know the past to predict
the future --- this is a natural assumption used in this and many other
applications. The system evolves by random transitions between states so that
the evolution of the probabilities is described by the so-called direct
Kolmogorov equation%
\begin{equation}
\frac{dP_{\mathbf{Y}^{\prime}}}{dt}=\sum_{\mathbf{Y}^{\prime\prime}}%
\overline{T}_{\mathbf{Y}^{\prime}\leftarrow\mathbf{Y}^{\prime\prime}%
}P_{\mathbf{Y}^{\prime\prime}}-\sum_{\mathbf{Y}^{\prime\prime}}\overline
{T}_{\mathbf{Y}^{\prime\prime}\leftarrow\mathbf{Y}^{\prime}}P_{\mathbf{Y}%
^{\prime}}, \label{Kolm1}%
\end{equation}
where $\overline{T}_{\mathbf{Y}^{\prime}\leftarrow\mathbf{Y}^{\prime\prime}}$
denotes the average transition rates from state $\mathbf{Y}^{\prime\prime}$ to
state $\mathbf{Y}^{\prime}$ and specify the transition coefficients of the
equation. The first term in (\ref{Kolm1}) evaluates all transitions into state
$\mathbf{Y}^{\prime}$ while the second term in (\ref{Kolm1}) sums up all
transitions from state $\mathbf{Y}^{\prime}$. These two terms can be assembled
into a single matrix $\overline{\overline{T}}_{\mathbf{Y}^{\prime}%
\mathbf{Y}^{\prime\prime}}$ so that
\begin{align}
\frac{dP_{\mathbf{Y}}}{dt}  &  =\overline{\overline{\mathbf{T}}}\mathbf{\cdot
}P_{\mathbf{Y}}=\sum_{\mathbf{Y}^{\prime\prime}}\overline{\overline{T}%
}_{\mathbf{Y}^{\prime}\mathbf{Y}^{\prime\prime}}P_{\mathbf{Y}^{\prime\prime}%
},\ \ \label{Kolm2}\\
\ \ \overline{\overline{T}}_{\mathbf{Y}^{\prime}\mathbf{Y}^{\prime\prime}}  &
=\overline{T}_{\mathbf{Y}^{\prime}\leftarrow\mathbf{Y}^{\prime\prime}}%
-\delta_{\mathbf{Y}^{\prime}\mathbf{Y}^{\prime\prime}}\sum_{\mathbf{Y}%
^{\prime\prime\prime}}\overline{T}_{\mathbf{Y}^{\prime\prime\prime}%
\leftarrow\mathbf{Y}^{\prime}}\ .
\end{align}
The matrix $\overline{\overline{T}}_{\mathbf{Y}^{\prime}\mathbf{Y}%
^{\prime\prime}}$ is conventionally called the transition rate matrix ---
operator $\overline{\overline{T}}_{\mathbf{Y}^{\prime}\mathbf{Y}^{\prime
\prime}}$ is specified in the following sections. The dimension of this
matrix, $3^{N}\times3^{N}$, is determined by the overall number of states and
only a small fraction of these values are non-zero. For the examples presented
in this work, $N$ is at least 500 and the full joint probability distribution
$P_{\mathbf{Y}}$ is represented by $3^{500}$ real numbers. Note that the value
$3^{500}$ exceeds by far the number of elementary particles in the known
universe (which is merely 10$^{80}$). It is needless to say that solving such
a large number of equations is completely impossible, even if we can
scrupulously specify all transition coefficients. Therefore, one needs to
consider possible simplifications.

\subsection{Marginal probabilities}

The problem becomes more traceable if expressed in terms of the marginal
probabilities
\begin{equation}
P^{(n)}=P\left(  Y_{i_{1}}^{\circ},Y_{i_{2}}^{\circ},...,Y_{i_{n}}^{\circ
}\right)  =\left\langle f^{(n)}\right\rangle =\left\langle \theta_{i_{1}%
}(Y_{i_{1}}^{\circ})\theta_{i_{2}}(Y_{i_{2}}^{\circ})...\theta_{i_{n}%
}(Y_{i_{n}}^{\circ})\right\rangle , \label{Pk}%
\end{equation}
where the $n\leq N$ and the set $i_{1},i_{2},...,i_{n}$ is a subset of length
$n$ of the overall set of nodes $1,2,...,N$. Note that $i_{1},i_{2},...,i_{n}$
is not a fixed particular set (say, the set of $1,2,...,n$) but reflect all
possible choices of $n$ elements from the full set $1,2,...,N$ of $N$
elements. Using ensemble averages in (\ref{Pk}) immediately tells us that the
value of $P^{(n)}$ does not depend on the order of the arguments, that is
$P\left(  Y_{i_{1}}^{\circ},Y_{i_{2}}^{\circ},...,Y_{i_{n}}^{\circ}\right)  $
is the same for any permutation of $Y_{i_{1}}^{\circ},Y_{i_{2}}^{\circ
},...,Y_{i_{n}}^{\circ}$. For example, $P(Y_{1}^{\circ},Y_{2}^{\circ}%
)=P(Y_{2}^{\circ},Y_{1}^{\circ}).$ The product $f^{(n)}=\theta_{i_{1}%
}(Y_{i_{1}}^{\circ})...\theta_{i_{n}}(Y_{i_{n}}^{\circ})$ is often called the
fine-grained distribution and its average is the corresponding probability
distribution $P^{(n)}=\left\langle f^{(n)}\right\rangle $. If $n=N$, then
$P^{(N)}=P_{Y}$ represents the full joint probability. Since, obviously,%
\[
\theta_{i}(Y^{\prime})\theta_{i}(Y^{\prime\prime})=\left\{
\begin{array}
[c]{cc}%
\theta_{i}(Y^{\prime}), & Y^{\prime}=Y^{\prime\prime}\\
0, & Y^{\prime}\neq Y^{\prime\prime}%
\end{array}
\right.
\]
the repeated nodes can be eliminated
\begin{equation}
P\left(  Y_{i_{1}}^{\circ},...,Y_{j}^{\prime},...,Y_{j}^{\prime\prime
}...,Y_{i_{n}}^{\circ}\right)  =P\left(  Y_{i_{1}}^{\circ},...,Y_{j}^{\prime
},...,Y_{i_{n}}^{\circ}\right)  \delta_{Y_{j}^{\prime}Y_{j}^{\prime\prime}}\ .
\label{Prep}%
\end{equation}
We do not need to consider any distributions with $n>N$ since the repeated
nodes can always be eliminated according to equation (\ref{Prep}).

The one-node (or first-order) probability distributions $P^{(1)}=P_{Y_{i}%
}=P\left(  Y_{i}^{\circ}\right)  =\left\langle \theta_{i}(Y_{i}^{\circ
})\right\rangle $ specify the probability of node $i$ being in a particular
state $Y_{i}^{\circ}$ (denoting one of S, I or R). These values, $P\left(
Y_{i}^{\circ}\right)  ,$ characterise only $3N$ states (i.e. $3$ states for
every node $i=1,...,N$). In the same way, the two-node probabilities
$P^{(2)}=P_{Y_{i}Y_{j}}=P\left(  Y_{i}^{\circ},Y_{j}^{\circ}\right)
=\left\langle \theta_{i}(Y_{i}^{\circ})\theta_{j}(Y_{j}^{\circ})\right\rangle
$ are characterised by $3N\times3N$ real values specifying probabilities for
all possible choices of nodes $i$ and $j$. When this cannot cause ambiguity,
the subscript is used to indicate the random variables that are characterised
by $P$. For example, $P_{Y_{i}Y_{j}}$\ is the marginal joint probability
distribution of $Y_{i}$ and $Y_{j},$\ and this distribution is a function of
two sample-space arguments $P\left(  Y_{i}^{\circ},Y_{j}^{\circ}\right)  $.
Note that any marginal probability $P^{(n)}$ can be conventionally expressed
in terms of the full joint probability $P_{\mathbf{Y}}$%
\begin{equation}
P\left(  Y_{1}^{\circ},Y_{2}^{\circ},...,Y_{n}^{\circ}\right)  =\int P\left(
Y_{1}^{\circ},Y_{2}^{\circ},...,Y_{n}^{\circ},Y_{n+1}^{\circ},...,Y_{N}%
^{\circ}\right)  dY_{n+1}^{\circ}...dY_{N}^{\circ}%
\end{equation}
or, similarly, in terms of marginal probabilities of $P^{(n^{\prime})}$ of a
higher order $n^{\prime}>n$.

\section{The governing equations}

\subsection{Equations for the fine-grained distributions}

Deriving equations for the full and marginal probabilities needs clear
notations and some care due to the large dimensions of the system under
consideration. It seems that following effective techniques introduced in
conditional methods \cite{KB99} and using fine-grained distributions is one of
the best possible choices. This approach is based on the following identity
\begin{equation}
\frac{df^{(n)}}{dt}=\frac{d\left(  \theta_{i_{1}}(Y_{i_{1}}^{\circ}%
)...\theta_{i_{n}}(Y_{i_{n}}^{\circ})\right)  }{dt}=\sum_{j=1}^{n}\left(
\theta_{i_{1}}(Y_{i_{1}}^{\circ})...\left[  \frac{d\theta_{i_{j}}(Y_{i_{j}%
}^{\circ})}{dt}\right]  ...\theta_{i_{n}}(Y_{i_{n}}^{\circ})\right)  ,
\label{df_dt}%
\end{equation}
which, after averaging, results in
\begin{equation}
\frac{dP^{(n)}}{dt}=\frac{d\left\langle f^{(n)}\right\rangle }{dt}%
=\left\langle \frac{df^{(n)}}{dt}\right\rangle =\sum_{j=1}^{n}\left\langle
\theta_{i_{1}}(Y_{i_{1}}^{\circ})...\left[  \frac{d\theta_{i_{j}}(Y_{i_{j}%
}^{\circ})}{dt}\right]  ...\theta_{i_{n}}(Y_{i_{n}}^{\circ})\right\rangle \ .
\label{dP_dt1}%
\end{equation}
Note that, since functions $\theta_{i}(...)$ takes discrete values $0$ and $1$
its conventional derivative does not exist and we customary imply generalised
derivatives. The full treatment of this problem is given in relevant textbooks
and Ref. \cite{KB99}, but we can simply use formal differentiation rules since
all these singularities disappear after averaging. We just need to evaluate
$d\theta_{i}/dt$ for the SIR model. This model is characterised by two
possible types of transitions --- infection $\Phi$ and recovery $\Psi$ --- so
that
\begin{equation}
\text{S}\overset{\Phi}{\longrightarrow}\text{I}\overset{\Psi}{\longrightarrow
}\text{R\ .} \label{S-I-R}%
\end{equation}
If $\Phi_{i}$ is an instance of infection of node $i$ from, say, node $j,$ and
$\Psi_{i}$ denotes an instance of recovery of node $i,$ then $\Phi_{i}$ and
$\Psi_{i}$ correspond to the following instantaneous transitions
\begin{equation}
\Phi_{i}=\sum_{j}T_{\text{I}_{i}\text{I}_{j}\leftarrow\text{S}_{i}\text{I}%
_{j}}=\sum_{j}p_{i}A_{ij}\theta_{i}(\text{S})\theta_{j}(\text{I}%
),\ \ \ \Psi_{i}=T_{\text{R}_{i}\leftarrow\text{I}_{i}}=q_{i}\theta
_{i}(\text{I}), \label{FiPsi}%
\end{equation}
where $A_{ij}$ is the adjacency matrix determining connectivity between the
nodes, $p_{i}$ specifies the probability of infection at node $i,$ and $q_{i}$
specifies the probability of recovery of this node. As indicated in equation
(\ref{FiPsi}), infection $T_{\text{I}_{i}\text{I}_{j}\leftarrow\text{S}%
_{i}\text{I}_{j}}$ is possible only when $Y_{i}=$S and $Y_{j}=$I while
recovery $T_{\text{R}_{i}\leftarrow\text{I}_{i}}$ requires that $Y_{i}=$I.
Note that transitions at the nodes $i$ and $j$ depend on $Y_{i}$ and $Y_{j},$
and do not directly depend on the states of the other nodes.

Equations (\ref{S-I-R}) and (\ref{FiPsi}) determine that
\begin{equation}
\frac{d\theta_{i}(\text{S})}{dt}=-\delta_{\phi}\Phi_{i},\ \ \ \frac
{d\theta_{i}(\text{I})}{dt}=\delta_{\phi}\Phi_{i}-\delta_{\psi}\Psi
_{i},\ \ \ \ \frac{d\theta_{i}(\text{R})}{dt}=\delta_{\psi}\Psi_{i}\ .
\label{dTet_dt}%
\end{equation}
Here, the delta-functions $\delta_{\phi}=\delta(t-t_{\phi})$ and $\delta
_{\psi}=\delta(t-t_{\psi})$ are\ used to indicate the presence of
singularities in the derivatives of the indicator functions $\theta($...$)$
pointing to jumps at random time moments: the instant of infection $t_{\phi}$
or the instant of recovery $t_{\psi}$. Equations (\ref{dTet_dt}) involve unit
jumps indicated by the Delta-functions and the rates of these jumps determined
by $\Phi_{i}$ and $\Psi_{i}$. For our purposes, the Delta-functions can simply
be omitted in all equations, since $\delta_{\phi}$ and $\delta_{\psi}$
disappear after averaging and do not affect the final equations --- we retain
these terms only for the sake of rigour. With the use of the following
indicator functions
\begin{equation}
\phi(Y^{\circ})=\left\{
\begin{array}
[c]{cc}%
-1, & Y^{\circ}=\text{S}\\
+1, & Y^{\circ}=\text{I}\\
0, & Y^{\circ}=\text{R}%
\end{array}
\right.  ,\ \ \ \psi(Y^{\circ})=\left\{
\begin{array}
[c]{cc}%
0, & Y^{\circ}=\text{S}\\
-1, & Y^{\circ}=\text{I}\\
+1, & Y^{\circ}=\text{R}%
\end{array}
\right.  ,
\end{equation}
equations (\ref{S-I-R})-(\ref{dTet_dt}) can be written as
\begin{equation}
\frac{d\theta_{i}(Y^{\circ})}{dt}=\delta_{\psi}q_{i}\psi(Y^{\circ})\theta
_{i}(\text{I})+\delta_{\phi}p_{i}\phi(Y^{\circ})\sum_{j}A_{ij}\theta
_{i}(\text{S})\theta_{j}(\text{I}) \label{dTet_dt2}%
\end{equation}
where $Y^{\circ}$ can take any of S, I or R. The substitution of
(\ref{dTet_dt2}) into (\ref{df_dt}) yields the evolution equation for the
fine-grained distribution
\begin{align}
\frac{df^{(n)}}{dt}  &  =\frac{d\theta_{i_{1}}(Y_{i_{1}}^{\circ}%
)...\theta_{i_{n}}(Y_{i_{n}}^{\circ})}{dt}\label{df_dt2}\\
&  =\sum_{j=1}^{n}\left(  \theta_{i_{1}}(Y_{i_{1}}^{\circ})...\left[
\delta_{\psi}q_{i_{j}}\psi(Y_{i_{j}}^{\circ})\theta_{i_{j}}(\text{I}%
)+\delta_{\phi}p_{i_{j}}\phi(Y_{i_{j}}^{\circ})\sum_{k}A_{i_{j}k}\theta
_{i_{j}}(\text{S})\theta_{k}(\text{I})\right]  ...\theta_{i_{n}}(Y_{i_{n}%
}^{\circ})\right) \nonumber
\end{align}

\subsection{Equations for marginal probabilities}

The governing equation for the marginal probabilities is the ensemble average
of equation (\ref{df_dt2})%
\begin{align}
&  \frac{dP\left(  Y_{i_{1}}^{\circ},Y_{i_{2}}^{\circ},...,Y_{i_{n}}^{\circ
}\right)  }{dt}\label{dP_dt-fiin}\\
&  =\sum_{j=1}^{n}\left[  q_{i_{j}}\psi(Y_{i_{j}}^{\circ})P\left(  Y_{i_{1}%
}^{\circ},...,\text{I}_{i_{j}},...,Y_{i_{n}}^{\circ}\right)  +p_{i_{j}}%
\phi(Y_{i_{j}}^{\circ})\sum_{i_{n+1}}A_{i_{j}i_{n+1}}P\left(  Y_{i_{1}}%
^{\circ},...,\text{S}_{i_{j}},...,Y_{i_{n}}^{\circ},\text{I}_{i_{n+1}}\right)
\right]  \ .\nonumber
\end{align}
Since this equation is quite general but cumbersome, we also give the first
and second order equations --- specific forms of equation (\ref{dP_dt-fiin})
for one-node $P^{(1)}=P_{Y_{i}}=P\left(  Y_{i}^{\circ}\right)  $ and two-node
$P^{(2)}=P_{Y_{i}Y_{j}}=P\left(  Y_{i}^{\circ},Y_{j}^{\circ}\right)  $
probability distributions.

\subsubsection{The first-order equations}

At the \textit{first order} we obtain
\begin{equation}
\frac{dP_{\text{S}_{i}}}{dt}=-\bar{\Phi}_{i},\ \ \ \ \frac{dP_{\text{I}_{i}}%
}{dt}=\bar{\Phi}_{i}-\bar{\Psi}_{i},\ \ \ \ \frac{dP_{\text{R}_{i}}}{dt}%
=\bar{\Psi}_{i}, \label{ord1-P}%
\end{equation}
where
\begin{equation}
\bar{\Phi}_{i}=\left\langle \Phi_{i}\right\rangle =p_{i}\sum_{j}%
A_{ji}P_{\text{I}_{j}\text{S}_{i}},\ \ \bar{\Psi}_{i}=\left\langle \Psi
_{i}\right\rangle =q_{i}P_{\text{I}_{i}} \label{ord1-Fi}%
\end{equation}
denote average rates of infection and recovery. These equations for one-node
probability distributions $P_{\text{S}_{i}},$ $P_{\text{I}_{i}}$ and
$P_{\text{R}_{i}}$\ also involve the two-node probability $P_{\text{I}%
_{j}\text{S}_{i}}=P_{\text{S}_{i}\text{I}_{j}}$.

\subsubsection{The second-order equations}

At the \textit{second order}, the equations for two-node probabilities are
obtained by substituting $n=2$ into (\ref{dP_dt-fiin}) or, equivalently, by
averaging (\ref{df_dt}) and (\ref{dTet_dt}) for $n=2$ and producing the
following set of equations
\begin{equation}%
\begin{array}
[c]{ccc}%
\frac{dP_{\text{S}_{i}\text{S}_{j}}}{dt}=-\bar{\Phi}_{i\text{S}_{j}}-\bar
{\Phi}_{j\text{S}_{i}}, & \frac{dP_{\text{I}_{i}\text{S}_{j}}}{dt}=\bar{\Phi
}_{i\text{S}_{j}}-\bar{\Phi}_{j\text{I}_{i}}-\bar{\Psi}_{i\text{S}_{j}}, &
\frac{dP_{\text{R}_{i}\text{S}_{j}}}{dt}=\bar{\Psi}_{i\text{S}_{j}}-\bar{\Phi
}_{j\text{R}_{i}},\\
\frac{dP_{\text{S}_{i}\text{I}_{j}}}{dt}=\bar{\Phi}_{j\text{S}_{i}}-\bar{\Phi
}_{i\text{I}_{j}}-\bar{\Psi}_{j\text{S}_{i}}, & \frac{dP_{\text{I}_{i}%
\text{I}_{j}}}{dt}=\bar{\Phi}_{i\text{I}_{j}}-\bar{\Psi}_{i\text{I}_{j}}%
+\bar{\Phi}_{j\text{I}_{i}}-\bar{\Psi}_{j\text{I}_{i}}, & \frac{dP_{\text{R}%
_{i}\text{I}_{j}}}{dt}=\bar{\Psi}_{i\text{I}_{j}}+\bar{\Phi}_{j\text{R}_{i}%
}-\bar{\Psi}_{j\text{R}_{i}},\\
\frac{dP_{\text{S}_{i}\text{R}_{j}}}{dt}=\bar{\Psi}_{j\text{S}_{i}}-\bar{\Phi
}_{i\text{R}_{j}}, & \frac{dP_{\text{I}_{i}\text{R}_{j}}}{dt}=\bar{\Psi
}_{j\text{I}_{i}}+\bar{\Phi}_{i\text{R}_{j}}-\bar{\Psi}_{i\text{R}_{j}}, &
\frac{dP_{\text{R}_{i}\text{R}_{j}}}{dt}=\bar{\Psi}_{i\text{R}_{j}}+\bar{\Psi
}_{j\text{R}_{i}},
\end{array}
\label{ord2-P}%
\end{equation}
where we denote $\bar{\Phi}_{iY_{j}}=\left\langle \Phi_{i}\theta
(Y_{j})\right\rangle \ $and$\ \bar{\Psi}_{iY_{j}}=\left\langle \Psi_{i}%
\theta(Y_{j})\right\rangle $ so that
\begin{equation}%
\begin{array}
[c]{ccc}%
\bar{\Phi}_{i\text{S}_{j}}=p_{i}\sum_{k\neq j}A_{ki}P_{\text{I}_{k}%
\text{S}_{i}\text{S}_{j}} & \bar{\Phi}_{i\text{I}_{j}}=p_{i}\sum_{k}%
A_{ki}P_{\text{I}_{k}\text{S}_{i}\text{I}_{j}} & \bar{\Phi}_{i\text{R}_{j}%
}=p_{i}\sum_{k\neq j}A_{ki}P_{\text{I}_{k}\text{S}_{i}\text{R}_{j}}\\
\bar{\Psi}_{i\text{S}_{j}}=q_{i}P_{\text{I}_{i}\text{S}_{j}} & \bar{\Psi
}_{i\text{I}_{j}}=q_{i}P_{\text{I}_{i}\text{I}_{j}} & \bar{\Psi}%
_{i\text{R}_{j}}=q_{i}P_{\text{I}_{i}\text{R}_{j}}%
\end{array}
\ . \label{ord2-Fi}%
\end{equation}
The matrix in (\ref{ord2-P}) is symmetric (that is $P(Y_{i}^{\prime}%
,Y_{j}^{\prime\prime})=P(Y_{j}^{\prime\prime},Y_{i}^{\prime})$ but, generally,
$P(Y_{i}^{\prime},Y_{j}^{\prime\prime})\neq P(Y_{i}^{\prime\prime}%
,Y_{j}^{\prime})$) involving only 6 independent equations. Since the one-node
probabilities can be obtained from two-node probabilities, equations
(\ref{ord1-P})-(\ref{ord1-Fi}) do not generally need to be solved in
conjunction with equations (\ref{ord2-P})-(\ref{ord2-Fi}). The second order
system, however, is not closed since the equations for two-node probabilities
involve the following three-node probabilities $P_{\text{S}_{i}\text{I}%
_{k}\text{S}_{j}},$ $P_{\text{S}_{i}\text{I}_{k}\text{I}_{j}}$ and
$P_{\text{S}_{i}\text{I}_{k}\text{R}_{j}}$. Some terms with $k=j$ are excluded
from the sums in (\ref{ord2-Fi}) since $P_{\text{S}_{i}\text{I}_{j}%
\text{S}_{j}}=P_{\text{S}_{i}\text{I}_{j}\text{R}_{j}}=0$ according to
(\ref{Prep}). While equations (\ref{ord2-P})-(\ref{ord2-Fi}) are generally
valid for any choice of nodes $i$ and $j$, we need to consider only the
connected nodes, i.e. those nodes $i$ and $j$ that ensure that $A_{ij}>0$.
Hence node $i$ is connected with both node $k$ and node $j$ in the three-node
probabilities $P_{\text{I}_{k}\text{S}_{i}Y_{j}}$\ that are summated in
(\ref{ord2-Fi}). The overall number of equations is of order of $\sim$3$^{2}E$
where $E$ is the number of edges in the graph.

\subsection{Conceptual interpretation of the governing equations}

One can note that the number of equations rapidly increases $\sim$3$^{n}$ with
the order of the system, but the system of equations remains unclosed. Indeed,
equation (\ref{dP_dt-fiin}) has the functional form of
\begin{equation}
\frac{dP^{(n)}}{dt}=\overline{\overline{\mathbf{T}}}_{\psi}^{(n)}\cdot
P^{(n)}+\overline{\overline{\mathbf{T}}}_{\phi}^{(n)}\cdot P^{(n+1)},
\label{dPnn1_dt}%
\end{equation}
so that the governing equations for $P^{(n)}$ involve $P^{(n+1)}$, while the
governing equations for $P^{(n+1)}$ involve $P^{(n+2)}$ and so on until $n=N$
is reached. Here, $\overline{\overline{\mathbf{T}}}_{\psi}^{(n)}$ and
$\overline{\overline{\mathbf{T}}}_{\phi}^{(n)}$ denote linear operators
(transitional matrices) that reflect transitions correspondingly associated
with recovery and infection; these operators are specified by the two terms on
the right-hand side of equation (\ref{dP_dt-fiin}). Since any probability
$P^{(N+1)}$ must have repeated nodes and, as noted in (\ref{Prep}), can be
expressed in terms of $P^{(N)},$ equation (\ref{dP_dt-fiin}) becomes
\begin{equation}
\frac{dP^{(N)}}{dt}=\overline{\overline{\mathbf{T}}}_{\psi}^{(N)}\cdot
P^{(N)}+\overline{\overline{\mathbf{T}}}_{\phi}^{(N)}\cdot P^{(N)}
\label{dPN_dt}%
\end{equation}
for $n=N.$ Unlike (\ref{dPnn1_dt}), this equation is closed and, of course,
coincide with the forward Kolmogorov equation (\ref{Kolm1}) that gives a
complete description for the whole system of $N$ nodes.

While equations (\ref{dPnn1_dt}) can be solved for small values of $n$, these
equations are not closed and force us to consider higher and higher orders
$n$. Equation (\ref{dPN_dt}) is closed but is practically unsolvable due to
its extremely large dimensionality. This is not accidental --- similar
problems are known to exist in large and complex systems including
multi-particle quantum mechanics and statistical physics. Equation
(\ref{dPN_dt}) is similar to the Liouville equation of statistical physics ---
both equations are exact and useless for simulations due to their extremely
large dimensionality. Equations (\ref{dPnn1_dt}) resemble the BBGKY
(Bogoliubov--Born--Green--Kirkwood--Yvon) hierarchy, which involve unclosed
equations \cite{bogoliubov1946kinetic,klimenko2009lagrangian}. The practical
way of solving such equations is in applying the hypothesis of molecular chaos
and decoupling distributions -- this procedure results in the Boltzmann
equation leading to the famous H-theorem. Similar problems can be found in
general particle modelling associated with reacting flows, producing a
hierarchy of equations of increasing dimensionality. At the systemic level,
there is a great deal of similarity between all these problems.

While we also use "chaotic decoupling" in this work, its application at the
first order as done in the conventional derivation of the Boltzmann equation
tends to produce inaccurate results. The systems we consider are not fully
chaotic and, as known from publications
\cite{klimenko2009lagrangian,KlimenkoPope2012CTM, klimenko2019evolution}, this
is the first sign of emerging complexity. In complex systems, interactions
between elements lead to substantial dependencies between them violating
"chaotic assumptions" and forcing us to consider multiparticle, multinode and
multivariable distributions.

\subsection{Monte-Carlo simulations}

If the evaluation of the probability distributions is difficult or impossible,
one of the common solutions is resorting to Monte-Carlo simulations, which
direct emulations of the underlying stochastic processes. Typically
Monte-Carlo simulations are more computationally expensive than low-order
distribution models but are very much affordable in comparison with solving
equations for full joint distributions. As with any modelling method,
Monte-Carlo simulations have their pluses and minuses. In the context of the
network SIR model, the Markov chain model is specified for a sufficiently
small time step $\Delta t$ and every node $i=1,...,N$ by the following
transitions
\begin{align}
\text{S}_{i}  &  \longrightarrow\text{I}_{i}\text{ with the probability
}p_{\Delta t}=p_{i}\Delta t\sum_{j}A_{ij}\theta_{j}(\text{I}),\\
\text{I}_{i}  &  \longrightarrow\text{R}_{i}\text{ with the probability
}q_{\Delta t}=q_{i}\Delta t\ .
\end{align}
The numerical issues are discussed further in the simulation section.

While using stochastic simulations, we still wish to obtain typical or average
characteristics and this may be problematic. First, since epidemics are
fundamentally unsteady processes, time averaging is not suitable for them. We,
however, may try to average over nodes, assuming that the network does not
have a strong localisation in the physical space associated with spatial
inhomogeneity. This averaging may work as long as values at different nodes
are not correlated, which, as noted above, is generally not correct. In the
present work, we combine averaging over nodes with ensemble averaging; that is
simulations are run independently many times and then average characteristics
are evaluated. This increases expenses associated with Monte-Carlo simulations.

There is another problem associated with stochastic simulations: real-world
systems may involve $\sim$10$^{6}$ elements (individuals) while we might use a
graph of $\sim$10$^{3}$ nodes to run simulations. The question of scaling up
is not trivial. One issue is preserving the node degree distribution (which
significantly affects simulations) when scaling networks --- this issue is
discussed further in the simulation section. The other issue is the
possibility of global and local extinctions, which, as we know from the
simulations of reacting flows, makes modelling complicated. Extinctions occur
when nodes (individuals) recover before transferring the infection. The case
of\ the reproduction number being close to unity is most complicated since the
process may or may not become extinct depending on realisations. Since each
node in simulations effectively represents a thousand individuals, it is clear
that extinctions between a few elements are more probable than among thousands
of individuals under the same conditions.

\section{Closures for marginal distributions}

\subsection{The first-order closure}

In this section, we conceptually follow Boltzmann's hypothesis of molecular
chaos that allows for the representation of two-particle distributions\ as the
product of the corresponding one-particle distributions. In the context of the
first-order system, which needs a closure for $P_{\text{I}_{j}\text{S}_{i}},$
this implies that
\begin{equation}
P_{\text{I}_{j}\text{S}_{i}}=\left\{
\begin{array}
[c]{cc}%
P_{\text{I}_{j}}P_{\text{S}_{i}}, & i\neq j\\
0, & i=j
\end{array}
\right.  \label{cls-1}%
\end{equation}
--- the two-node distribution $P_{\text{I}_{j}\text{S}_{i}}$ is assumed to be
a product of the one-node distributions $P_{\text{I}_{j}}$ and $P_{\text{S}%
_{i}}$ implementing simple unconditional decoupling. Note that $P_{\text{I}%
_{i}\text{S}_{i}}$ does not enter equation (\ref{ord1-Fi}) since $A_{ii}=0$
and does not need to be specified; therefore, assuming $P_{\text{I}%
_{j}\text{S}_{i}}=P_{\text{I}_{j}}P_{\text{S}_{i}}$ for all $i$ and $j$ yields
exactly the same model. For the sake of simplicity, the approximation details
that do not affect the model are omitted from further consideration.

Substitution of the first-order decoupling closure (\ref{cls-1}) into
equations (\ref{ord1-P})-(\ref{ord1-Fi}) results in the closed system for
one-node probability distributions:%
\begin{equation}
\frac{dP_{\text{S}_{i}}}{dt}=-\bar{\Phi}_{i},\ \ \ \ \frac{dP_{\text{I}_{i}}%
}{dt}=\bar{\Phi}_{i}-\bar{\Psi}_{i},\ \ \ \ \frac{dP_{\text{R}_{i}}}{dt}%
=\bar{\Psi}_{i}, \label{cls1-P}%
\end{equation}%
\begin{equation}
\bar{\Phi}_{i}=\sum_{j}p_{i}A_{ij}P_{\text{I}_{j}}P_{\text{S}_{i}%
},\ \ \ \ \bar{\Psi}_{i}=q_{i}P_{\text{I}_{i}}, \label{cls1-Fi}%
\end{equation}
where $i=1,...,N$. The first-order model involves only $3N$ ordinary
differential equations but, as shown in the following sections, the
first-order decoupling is not particularly accurate due to stochastic
dependencies between neighbouring nodes.

\subsection{The ergodic closure}

This closure is suitable for the case when the adjacency matrix is decomposed
into two terms $A_{ij}=A_{ij}^{\circ}+A_{ij}^{\prime}$ so that the principal
term has relatively few significant connections $A_{ij}^{\circ}\sim1,$ while
the second term reflects the possibility of numerous but weak (or occasional)
connections $A_{ij}^{\prime}\ll1$. The second term either can be negligible or
may contribute to the overall evolution of the epidemic despite $A_{ij}%
^{\prime}\ll1$ due to a large number of the possible contacts. This
contribution can be evaluated assuming that $A_{ij}^{\prime}=\varepsilon\ll1$
is the same for all nodes (since its effect is averaged over a very large
number of possible contacts) leading to equation (\ref{ord1-Fi}) taking the
following form
\begin{equation}
\bar{\Phi}_{i}=p_{i}\sum_{j}A_{ji}^{\circ}P_{\text{I}_{j}\text{S}_{i}}%
+p_{i}\varepsilon N\bar{P}_{\text{IS}_{i}},\ \ \ \ \bar{P}_{\text{IS}_{i}%
}=\frac{1}{N}\sum_{j}P_{\text{I}_{j}\text{S}_{i}}\ ,
\end{equation}
The first term is subject to the first- and second-order closures discussed in
this section, while the node-average probability is evaluated as
\begin{equation}
\bar{P}_{\text{IS}_{i}}=\frac{1}{N}\sum_{j}\left\langle \theta_{j}%
(\text{I})\theta_{i}(\text{S})\right\rangle =\left\langle \theta_{i}%
(\text{S})P_{\text{I}}\right\rangle =P_{\text{S}_{i}}P_{\text{I}}\ ,
\label{cls15-Fi}%
\end{equation}
where the ergodic hypothesis%
\begin{equation}
\theta(\text{I})\overset{\text{def}}{=}\frac{1}{N}\sum_{j}\theta_{j}%
(\text{I})\approx P_{\text{I}}\overset{\text{def}}{=}\frac{1}{N}\sum
_{j}P_{\text{I}_{j}}%
\end{equation}
is applied, implying that the average over all nodes coincides with the
corresponding ensemble average. While decoupling (\ref{cls1-Fi}) may or may
not be accurate when applied to the principal part of the graph $A_{ij}%
^{\circ}$ requiring higher-order closures, decoupling (\ref{cls15-Fi}) applied
to secondary connections is much better since the states of weakly connected
nodes are not likely to be strongly correlated. Yet the ergodic hypothesis is
not exact, especially when extinctions are present. Indeed, by definition
$\theta($I$)=0$ for extinct realisations, while $P_{\text{I}}>0$ when some of
the realisations are not extinct. As in modelling of reacting flows,
extinctions tend to increase systemic complexity.

In the simulations presented in this work, we do not consider secondary
connections, assuming that $A_{ij}=1$ for connected nodes but, in the real
world, occasional transmissions which have very low probability for given $i$
and $j$ may contribute significantly when the population is large $N\gg1$.

\subsection{The second-order direct decoupling closure}

The second order closure implies that $P_{\text{S}_{i}\text{I}_{j}}$ is not
approximated by (\ref{cls-1}) but modelled using equations (\ref{ord2-P}%
)-(\ref{ord2-Fi}). The\ second-order equation for $P_{\text{I}_{j}\text{S}%
_{i}}$
\begin{equation}
\frac{dP_{\text{I}_{j}\text{S}_{i}}}{dt}=-p_{i}\sum_{k}A_{ki}P_{\text{I}%
_{k}\text{S}_{i}\text{I}_{j}}+p_{j}\sum_{k}A_{kj}P_{\text{I}_{k}\text{S}%
_{j}\text{S}_{i}}-q_{j}P_{\text{I}_{j}\text{S}_{i}} \label{cls2-eq}%
\end{equation}
remains unclosed due to presence of the three-node probabilities
$P_{\text{I}_{k}\text{S}_{i}\text{I}_{j}}$ and $P_{\text{I}_{k}\text{S}%
_{j}\text{S}_{i}}$, which need to be approximated. Note that although equation
(\ref{cls2-eq}) is valid for any $i,j$ $\in1,...,N$, we need evaluation of
$P_{\text{I}_{j}\text{S}_{i}}$ only when $A_{ji}>0,$ i.e. for distinct
connected nodes $i$ and $j$. The following unconditional approximations%
\begin{equation}
P_{\text{I}_{k}\text{S}_{i}\text{I}_{j}}=\left\{
\begin{array}
[c]{cc}%
P_{\text{I}_{k}\text{S}_{i}}P_{\text{I}_{j}}, & k\neq j\\
P_{\text{I}_{j}\text{S}_{i}}, & k=j
\end{array}
\right.  ,\ \ \ \ P_{\text{I}_{k}\text{S}_{j}\text{S}_{i}}=\left\{
\begin{array}
[c]{cc}%
P_{\text{I}_{k}\text{S}_{j}}P_{\text{S}_{i}}, & k\neq i\\
0, & k=i
\end{array}
\right.  \label{cls-2}%
\end{equation}
lead to the system
\begin{equation}
\frac{dP_{\text{S}_{i}}}{dt}=-\bar{\Phi}_{i},\ \ \ \ \frac{dP_{\text{I}_{i}}%
}{dt}=\bar{\Phi}_{i}-\bar{\Psi}_{i},\ \ \ \ \frac{dP_{\text{R}_{i}}}{dt}%
=\bar{\Psi}_{i}\ , \label{dP_Y_dt2}%
\end{equation}%
\begin{equation}
\bar{\Phi}_{i}=p_{i}\sum_{j}A_{ij}P_{\text{I}_{j}\text{S}_{i}},\ \ \ \ \bar
{\Psi}_{i}=q_{i}P_{\text{I}_{i}}\ ,
\end{equation}

\begin{equation}
\frac{dP_{\text{I}_{j}\text{S}_{i}}}{dt}=-p_{i}\sum_{k}A_{ki}P_{\text{I}%
_{k}\text{S}_{i}}P_{\text{I}_{j}}-\underset{\text{(a)}}{\underbrace{p_{i}%
A_{ji}P_{\text{I}_{j}\text{S}_{i}}\left(  1-P_{\text{I}_{j}}\right)  }}%
+p_{j}\sum_{k}A_{kj}P_{\text{I}_{k}\text{S}_{j}}P_{\text{S}_{i}}%
-\underset{\text{(b)}}{\underbrace{p_{j}A_{ij}P_{\text{I}_{i}\text{S}_{j}%
}P_{\text{S}_{i}}}}-q_{j}P_{\text{I}_{j}\text{S}_{i}}\ , \label{dP_IS_dt2}%
\end{equation}
which is closed system of $4N$ differential equations.

If more simple closures $P_{\text{I}_{k}\text{S}_{i}\text{I}_{j}}%
=P_{\text{I}_{k}\text{S}_{i}}P_{\text{I}_{j}}$ and $P_{\text{I}_{k}%
\text{S}_{j}\text{S}_{i}}=P_{\text{I}_{k}\text{S}_{j}}P_{\text{S}_{i}}$\ for
all $i,j,k$\ are used instead of (\ref{cls-2}), then terms (a) and (b) vanish
from equation (\ref{dP_IS_dt2}). These simple closures are obviously incorrect
since $P_{\text{I}_{j}\text{S}_{i}\text{I}_{j}}=P_{\text{I}_{j}\text{S}_{i}%
}\neq P_{\text{I}_{j}\text{S}_{i}}P_{\text{I}_{j}}$ and $P_{\text{I}%
_{i}\text{S}_{j}\text{S}_{i}}=0\neq P_{\text{I}_{i}\text{S}_{j}}%
P_{\text{S}_{i}}$ according to (\ref{Prep}). Equation (\ref{dP_IS_dt2}) can be
compared with the equation
\begin{equation}
\frac{dP_{\text{I}_{j}}P_{\text{S}_{i}}}{dt}=-p\sum_{k}A_{ki}P_{\text{I}%
_{k}\text{S}_{i}}P_{\text{I}_{j}}+p\sum_{k}A_{kj}P_{\text{I}_{k}\text{S}_{j}%
}P_{\text{S}_{i}}-qP_{\text{I}_{j}}P_{\text{S}_{i}} \label{dP_IS_dt3}%
\end{equation}
for the product $P_{\text{I}_{j}}P_{\text{S}_{i}}$ obtained from
(\ref{ord1-P})-(\ref{ord1-Fi}). It is easy to see that equation
(\ref{dP_IS_dt3}) coincides with equation (\ref{dP_IS_dt2}) whenever terms (a)
and (b) are removed. This implies that, without effects of terms (a) and (b),
$P_{\text{I}_{j}\text{S}_{i}}=P_{\text{I}_{j}}P_{\text{S}_{i}}$ and the
second-order model is functionally reduced to the first order.

\subsection{The second-order conditional closure}

In addition to equation (\ref{cls2-eq}), this \ closure uses another
second-order equation \
\begin{equation}
\frac{dP_{\text{S}_{j}\text{S}_{i}}}{dt}=-p_{j}\sum_{k}A_{kj}P_{\text{I}%
_{k}\text{S}_{j}\text{S}_{i}}-p_{i}\sum_{k}A_{ki}P_{\text{I}_{k}\text{S}%
_{i}\text{S}_{j}} \label{cls3-eq}%
\end{equation}
for the two-node joint probability $P_{\text{S}_{j}\text{S}_{i}}$ obtained in
(\ref{ord2-P})-(\ref{ord2-Fi}). Both equations (\ref{cls2-eq}) and
(\ref{cls3-eq}) need closures for three-node probabilities $P_{\text{I}%
_{k}\text{S}_{j}\text{S}_{i}}$and $P_{\text{I}_{k}\text{S}_{i}\text{S}_{j}}$,
which is based on the following transformations $P_{\text{I}_{k}\text{S}%
_{i}\text{I}_{j}}=P_{\text{I}_{k}\text{I}_{j}|\text{S}_{i}}P_{\text{S}_{i}}$
and $P_{\text{I}_{k}\text{S}_{i}\text{S}_{j}}=P_{\text{I}_{k}\text{S}%
_{j}|\text{S}_{i}}P_{\text{S}_{i}}$ where vertical bar denotes conditional
probabilities, for example the probability $P_{\text{I}_{k}\text{S}%
_{j}|\text{S}_{i}}=P($I$_{k},$S$_{j}|$S$_{i})$ is conditioned on $Y_{i}%
=$S$_{i}$. Note that it is the central node $i$ which is connected by the
graph to its nebouring nodes $j$ and $k$ (so that $A_{ij}>0$ and $A_{ik}>0)$
that is selected for conditioning. The conditional closure is based on the
following decoupling
\begin{equation}
P_{\text{I}_{k}\text{S}_{j}|\text{S}_{i}}=P_{\text{I}_{k}|\text{S}_{i}%
}P_{\text{S}_{j}|\text{S}_{i}}\text{ \ \ and \ \ }P_{\text{I}_{k}\text{I}%
_{j}|\text{S}_{i}}=P_{\text{I}_{k}|\text{S}_{i}}P_{\text{I}_{j}|\text{S}_{i}%
}\ . \label{cls3dcp}%
\end{equation}
While these relations are approximate, it is well-known that conditional
decoupling implemented in conditional methods (e.g. Conditional Moment Closure
and Multiple Mapping Conditioning --- effective models used in simulations of
reacting flows) is much better than any analogous unconditional decoupling. We
also need to note the following identities
\begin{equation}
P_{\text{I}_{j}\text{I}_{j}|\text{S}_{i}}=P_{\text{I}_{j}|\text{S}_{i}}\text{
\ \ and \ \ }P_{\text{I}_{j}\text{S}_{j}|\text{S}_{i}}=0\ , \label{cls3id}%
\end{equation}
and obtain the relations%
\begin{equation}
P_{\text{I}_{k}\text{S}_{i}\text{I}_{j}}=\left\{
\begin{array}
[c]{cc}%
P_{\text{I}_{k}|\text{S}_{i}}P_{\text{S}_{i}}P_{\text{I}_{j}|\text{S}_{i}}, &
k\neq j\\
P_{\text{I}_{j}\text{S}_{i}}, & k=j
\end{array}
\right.  ,\ \ \ \ P_{\text{I}_{k}\text{S}_{i}\text{S}_{j}}=\left\{
\begin{array}
[c]{cc}%
P_{\text{I}_{k}|\text{S}_{i}}P_{\text{S}_{i}}P_{\text{S}_{j}|\text{S}_{i}} &
k\neq j\\
0, & k=j
\end{array}
\right.  \ ,
\end{equation}
which consistently implement conditional decoupling. The conditional closure
results in the following system of equations
\begin{equation}
\frac{dP_{\text{S}_{i}}}{dt}=-\bar{\Phi}_{i},\ \ \ \ \frac{dP_{\text{I}_{i}}%
}{dt}=\bar{\Phi}_{i}-\bar{\Psi}_{i},\ \ \ \ \frac{dP_{\text{R}_{i}}}{dt}%
=\bar{\Psi}_{i}, \label{cls3e1}%
\end{equation}%
\begin{equation}
\frac{dP_{\text{S}_{j}\text{S}_{i}}}{dt}=-p_{j}\sum_{k}A_{kj}P_{\text{I}%
_{k}\text{S}_{j}}P_{\text{S}_{i}|\text{S}_{j}}+\underset{\text{(c)}%
}{\underbrace{p_{j}A_{ij}P_{\text{I}_{i}\text{S}_{j}}P_{\text{S}_{i}%
|\text{S}_{j}}}}-p_{i}\sum_{k}A_{ki}P_{\text{I}_{k}\text{S}_{i}}%
P_{\text{S}_{j}|\text{S}_{i}}+\underset{\text{(d)}}{\underbrace{p_{i}%
A_{ji}P_{\text{I}_{j}\text{S}_{i}}P_{\text{S}_{j}|\text{S}_{i}}}},
\label{cls3e2}%
\end{equation}%
\begin{align}
\frac{dP_{\text{I}_{j}\text{S}_{i}}}{dt}  &  =-p_{i}\sum_{k}A_{ki}%
P_{\text{I}_{k}\text{S}_{i}}P_{\text{I}_{j}|\text{S}_{i}}\label{cls3e3}\\
&  -\underset{\text{(a)}}{\underbrace{p_{i}A_{ji}P_{\text{I}_{j}\text{S}_{i}%
}\left(  1-P_{\text{I}_{j}|\text{S}_{i}}\right)  }}+p_{j}\sum_{k}%
A_{kj}P_{\text{I}_{k}\text{S}_{j}}P_{\text{S}_{i}|\text{S}_{j}}%
-\underset{\text{(b)}}{\underbrace{p_{j}A_{ij}P_{\text{I}_{i}\text{S}_{j}%
}P_{\text{S}_{i}|\text{S}_{j}}}}-q_{j}P_{\text{I}_{j}\text{S}_{i}},\nonumber
\end{align}
where
\begin{equation}
\bar{\Phi}_{i}=p_{i}\sum_{j}A_{ij}P_{\text{I}_{j}\text{S}_{i}},\ \ \ \ \bar
{\Psi}_{i}=q_{i}P_{\text{I}_{i}},\ \ \ P_{\text{S}_{j}|\text{S}_{i}}%
=\frac{P_{\text{S}_{j}\text{S}_{i}}}{P_{\text{S}_{i}}},\ \ \ \ P_{\text{I}%
_{j}|\text{S}_{i}}=\frac{P_{\text{I}_{j}\text{S}_{i}}}{P_{\text{S}_{i}}}\ .
\label{cls3e4}%
\end{equation}
The system involves $5N$ ordinary differential equations\ and represents a
closed second-order model based on conditional decoupling analogous to those
used in conditional methods.

As in the previous subsection, overriding conditional identities
(\ref{cls3id}) by (\ref{cls3dcp}) removes the terms (a), (b), (c) and (d) in
equations (\ref{cls3e2}) and (\ref{cls3e4}), which makes these equations
coincident with the following identities
\begin{equation}
\frac{dP_{\text{S}_{j}}P_{\text{S}_{i}}}{dt}=-p_{j}\sum_{k}A_{kj}%
P_{\text{I}_{k}\text{S}_{j}}P_{\text{S}_{i}}-p_{i}\sum_{k}A_{ki}%
P_{\text{I}_{k}\text{S}_{i}}P_{\text{S}_{j}},
\end{equation}%
\begin{equation}
\frac{dP_{\text{I}_{j}}P_{\text{S}_{i}}}{dt}=-p_{i}\sum_{k}A_{ki}%
P_{\text{I}_{k}\text{S}_{i}}P_{\text{I}_{j}}+p_{j}\sum_{k}A_{kj}%
P_{\text{I}_{k}\text{S}_{j}}P_{\text{S}_{i}}-q_{j}P_{\text{I}_{j}}%
P_{\text{S}_{i}}\ .
\end{equation}
This effectively leads to equalities $P_{\text{S}_{j}\text{S}_{i}}%
=P_{\text{S}_{j}}P_{\text{S}_{i}}$ and $P_{\text{I}_{j}\text{S}_{i}%
}=P_{\text{I}_{j}}P_{\text{S}_{i}}$, functionally reducing the conditional
second-order closure to the first order.

\section{Propagation of epidemic on simple graphs.}

This section investigates propagation of SIR epidemic on relatively simple
algorithmically generated graphs allowing for exact solutions. These results
are subsequently compared with the closures.

\begin{figure}
\begin{center}
\includegraphics[width=.75\linewidth,page=1,trim=6cm 8cm 6.2cm 9cm, clip ]{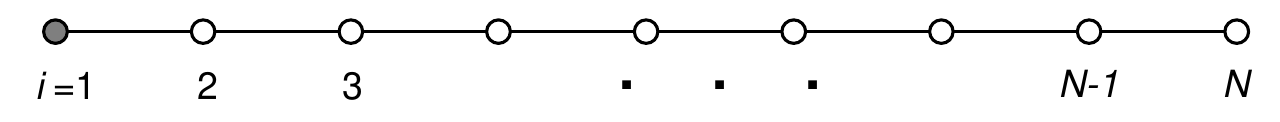}

\caption{One-dimensional connected graph with initial infection of the first node $i=1$. \label{fig0}}

\end{center}
\end{figure}

\subsection{Exact solution in one-dimensional case}

First, we examine the case\ of one-dimensional propagation of infection, which
allows for a relatively simple analytical solution. This is very much
analogous to the one-dimensional interpretation used in the original Ising
model. Only one connected graph is possible in one dimension that connects the
nodes $[1,2],$ $[2,3],...,[i,i=1],...$ that is $A_{i,i\pm1}=1$\ as shown in
Figure \ref{fig0}.\ The initial conditions are specified by
\begin{equation}
Y_{1}=\text{I},\ \ \ Y_{i}=\text{S \ for }t=0\text{\ and\ }i=2,3,...,N
\end{equation}
with a sufficiently large $N$. \ The probabilities of infection $p$ and
recovery $q$ are deemed to be node-independent constants that is $p_{i}=p$ and
$q_{i}=q$ for all $i$.

We use $P_{i\text{I}}$ and $P_{i\text{R}}$ to denote the following marginal
probabilities
\begin{equation}
P_{i\text{I}}=P\left(  \text{I}_{i},\text{S}_{i+1},...,\text{S}_{N}\right)
\text{\ \ and\ \ }P_{i\text{R}}=P\left(  \text{R}_{i},\text{S}_{i+1}%
,...,\text{S}_{N}\right)  \ ,\text{\ }%
\end{equation}
where symbols $i$I and $i$R are used as abbreviated notations for the
corresponding states
\begin{equation}
i\text{I}\overset{\text{def}}{=}\left[  Y_{1},...,Y_{i-1},\text{I}%
_{i},\text{S}_{i+1},...,\text{S}_{N}\right]  \text{ \ \ and \ }i\text{R}%
\overset{\text{def}}{=}\left[  Y_{1},...,Y_{i-1},\text{R}_{i},\text{S}%
_{i+1},...,\text{S}_{N}\right]  \ .
\end{equation}
Note that $Y_{1},...Y_{i-1}$ can be either I or R. These states are subject to
the transitions
\begin{equation}
\overline{T}_{(i+1)\text{I}\leftarrow i\text{I}}=pP_{i\text{I}}%
,\ \ \ \overline{T}_{i\text{R}\leftarrow i\text{I}}=qP_{i\text{I}}%
\end{equation}
supplemented by other transitions involving changes in $Y_{1},...,Y_{i-1}$,
which do not need to be considered. The governing equation for probability
takes the form
\begin{equation}
\frac{dP_{i\text{I}}}{dt}=pP_{(i-1)\text{I}}-\left(  p+q\right)  P_{i\text{I}%
},\ \ \ \frac{dP_{i\text{R}}}{dt}=qP_{i\text{I}}, \label{D1P2}%
\end{equation}
where $i=1,...,N$ and we formally put $P_{0\text{I}}=0$. The one-node
probabilities can be easily evaluated from
\begin{equation}
\frac{dP_{\text{S}_{i}}}{dt}=-pP_{(i-1)\text{I}},\ \ \ \frac{dP_{\text{I}_{i}%
}}{dt}=pP_{(i-1)\text{I}}-qP_{\text{I}_{i}},\ \ \ \frac{dP_{\text{R}_{i}}}%
{dt}=qP_{\text{I}_{i}}\ . \label{D1P1}%
\end{equation}

\subsection{Comparison with the closures}

For one-dimensional lattice considered here, equations (\ref{cls1-P}%
)-(\ref{cls1-Fi}), which are associated with the first-order closure, take the
form
\begin{equation}
\frac{dP_{\text{S}_{i}}}{dt}=-pP_{\text{I}_{i-1}}P_{\text{S}_{i}%
},\ \ \ \ \frac{dP_{\text{I}_{i}}}{dt}=pP_{\text{I}_{i-1}}P_{\text{S}_{i}%
}-qP_{\text{I}_{i}},\ \ \ \ \frac{dP_{\text{R}_{i}}}{dt}=qP_{\text{I}_{i}}\ .
\end{equation}
These equations are quite different from the exact equations (\ref{D1P2}%
)-(\ref{D1P1}).

For the second-order direct decoupling closure (\ref{dP_Y_dt2}%
)-(\ref{dP_IS_dt2}), the equations for one-node probabilities coincide with
(\ref{D1P1}), assuming $P_{\text{I}_{i}\text{S}_{i+1}}=P_{i\text{I}}$ and
$P_{\text{I}_{i}\text{S}_{i-1}}=0$. The closure equation for the two-node
probability $P_{\text{I}_{i}\text{S}_{i+1}}$
\begin{equation}
\frac{dP_{\text{I}_{i}\text{S}_{i+1}}}{dt}=pP_{\text{I}_{i-1}\text{S}_{i}%
}P_{\text{S}_{i+1}}-\left(  p+q\right)  P_{\text{I}_{i}\text{S}_{i+1}}
\label{D1cls2}%
\end{equation}
is nevertheless different from (\ref{D1P2}) due to the presence of an
additional multiplier, $P_{\text{S}_{i+1}},$ in the first term on the
right-hand side of equation (\ref{D1cls2}).

The second-order conditional closure (\ref{cls3e1})-(\ref{cls3e4}) also
reproduces (\ref{D1P1}), assuming $P_{\text{I}_{i}\text{S}_{i+1}}%
=P_{i\text{I}}$ and $P_{\text{I}_{i}\text{S}_{i-1}}=0,$ while the closure
equation for the two-node probability
\begin{equation}
\frac{dP_{i}}{dt}=+pP_{i-1}P_{\text{S}_{i+1}|\text{S}_{i}}-\left(  p+q\right)
P_{i}%
\end{equation}
is functionally the same as the exact equation (\ref{D1P2}) since
$P_{\text{S}_{i+1}|\text{S}_{i}}=1$ under these conditions.

\begin{figure}
\begin{center}
\includegraphics[width=.75\linewidth,page=1,trim=1.2cm 0.8cm 2.4cm 13cm, clip ]{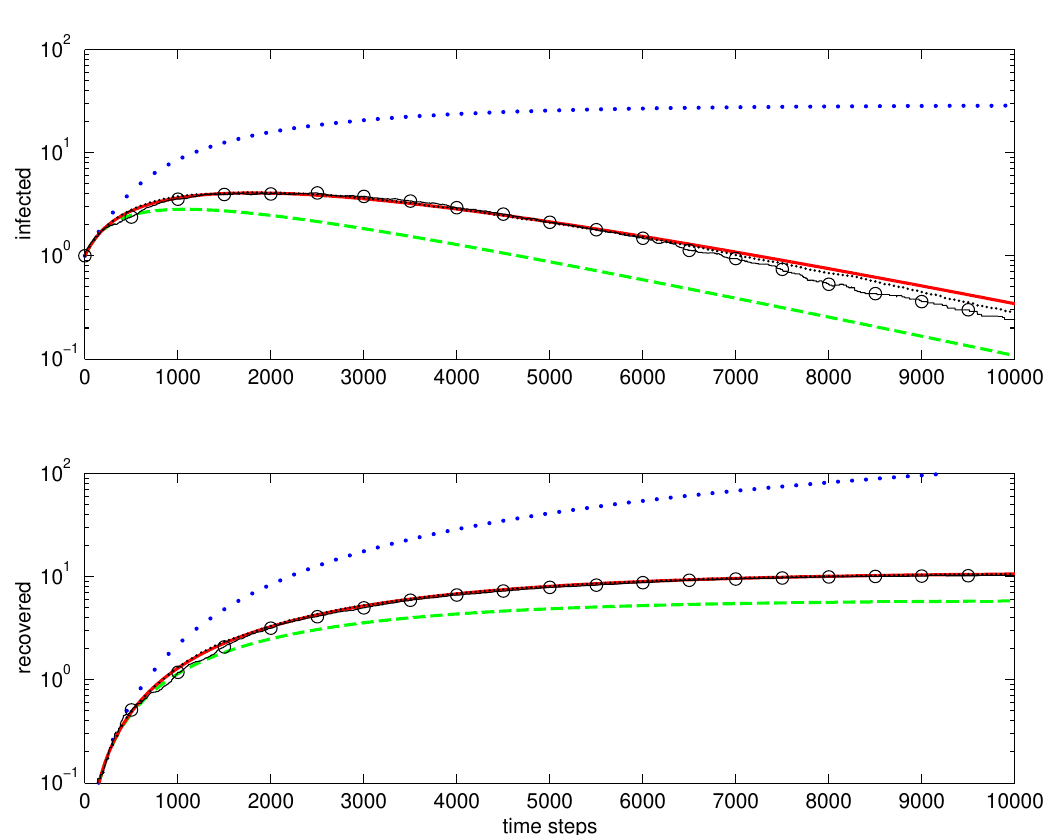}

\caption{Modelling epidemic in one-dimensional case: total infected (top figure) and recovered (bottom figure). 
Lines:    
$\bullet ~\bullet ~\bullet ~\bullet $ the first-order closure;
-- -- -- --    the second-order direct decoupling closure;
---------      the second-order conditional closure;
o---o---o      Monte-Carlo, ensemble averaging over 100 realisations;
$\cdot\cdot\cdot\cdot\cdot$ Monte-Carlo, ensemble averaging over 1000 realisations. 
Simulation parameters: $\tilde{p}=p\Delta t=5\times 10^{-3},$  $\tilde{q}=q\Delta t=8\times 10^{-4}$. 
\label{fig1}}

\end{center}
\end{figure}

An example of one-dimensional simulations is shown in Figure \ref{fig1}. The
first-order closure does not reproduce correct behaviour. The second-order
direct decoupling closure is qualitatively correct but overestimates
extinctions. The second-order conditional closure is accurate. Note that, as
the average number of infected nodes drops well below unity, averaging becomes
difficult for Monte-Carlo simulations, since most stochastic realisations do
not have any infected nodes.

\subsection{Epidemic propagation on a tree.}

The first infected node is assigned number $1$ each other node is
characterised by its number $i$ and the distance $l_{i}$ from node 1, which is
called level. Obviously, $l_{1}=0$. In a tree any connected nodes belong to
neighbouring levels, that is nodes $i$ and $j$ can be connected only if
$l_{j}=l_{i}\pm1$. The evolution equations for the marginal probabilities can
be obtained from the first (\ref{ord1-P})-(\ref{ord1-Fi}) and second
(\ref{ord2-P})-(\ref{ord2-Fi}) order equations by taking into account that the
graph under consideration is a tree.

Consider three-node probabilities $P_{\text{I}_{k}\text{S}_{i}Y_{j}}=P($%
I$_{k},$S$_{i},Y_{j})$\ used in (\ref{ord2-Fi}). Since node $i$ is connected
to nodes $k$ and $j$, there are only two possibilities $P($I$_{k}^{l-1}%
,$S$_{i}^{l},Y_{j}^{l+1})$ and $P($I$_{k}^{l+1},$S$_{i}^{l},Y_{j}^{l+1})$ for
these probabilities where $Y_{j}^{l}$ indicates state $Y$ of node $j$ that
belongs to level $l$. It is easy to see that
\begin{equation}
P(\text{I}_{k}^{l-1},\text{S}_{i}^{l},\text{S}_{j}^{l+1})\geq0\text{,
\ \ \ }P(\text{I}_{k}^{l-1},\text{S}_{i}^{l},\text{I}_{j}^{l+1})=P(\text{I}%
_{k}^{l-1},\text{S}_{i}^{l},\text{R}_{j}^{l+1})=0
\end{equation}
and
\begin{equation}
P(\text{I}_{k}^{l+1},\text{S}_{i}^{l},\text{S}_{j}^{l+1})=P(\text{I}_{k}%
^{l+1},\text{S}_{i}^{l},\text{I}_{j}^{l+1})=P(\text{I}_{k}^{l+1},\text{S}%
_{i}^{l},\text{R}_{j}^{l+1})=0\ .
\end{equation}
This implies that all three-node correlations of interest are zeros with
exception of $P($I$_{k}^{l-1},$S$_{i}^{l},$S$_{j}^{l+1}),$ which can be
expressed as
\begin{equation}
P(\text{I}_{k}^{l-1},\text{S}_{i}^{l},\text{S}_{j}^{l+1})=P(\text{I}_{k}%
^{l-1},\text{S}_{i}^{l})\ ,
\end{equation}
since I$_{k}^{l-1}$ and S$_{i}^{l}$ always imply $Y_{j}^{l+1}=$S when nodes
$i\ $and $j$ are connected. Substitution of these equalities into
(\ref{ord1-P})-(\ref{ord2-Fi}) results in the following system
\begin{equation}
\frac{dP(\text{S}_{i}^{l})}{dt}=-\bar{\Phi}_{i},\ \ \ \ \frac{dP(\text{I}%
_{i}^{l})}{dt}=\bar{\Phi}_{i}-\bar{\Psi}_{i},\ \ \ \ \frac{dP(\text{R}_{i}%
^{l})}{dt}=\bar{\Psi}_{i},
\end{equation}%
\begin{equation}
\bar{\Phi}_{i}=\left\langle \Phi_{i}\right\rangle =p_{i}\sum_{k}%
A_{ji}P(\text{I}_{j}^{l-1},\text{S}_{i}^{l}),\ \ \ \bar{\Psi}_{i}=\left\langle
\Psi_{i}\right\rangle =q_{i}P(\text{I}_{i}^{l}),
\end{equation}%
\begin{equation}
\frac{dP(\text{I}_{j}^{l},\text{S}_{i}^{l+1})}{dt}=p_{i}\sum_{k}%
A_{ki}P(\text{I}_{k}^{l-1},\text{S}_{j}^{l})-q_{i}P(\text{I}_{j}^{l}%
,\text{S}_{i}^{l+1})\ .
\end{equation}
This system of equations is closed and does not need any further assumptions.
Note that the same equations can be derived from the second-order conditional closure.

\begin{figure}
\begin{center}
\includegraphics[width=.9\linewidth,page=1,trim=1cm 0.5cm 2cm 13.5cm, clip ]{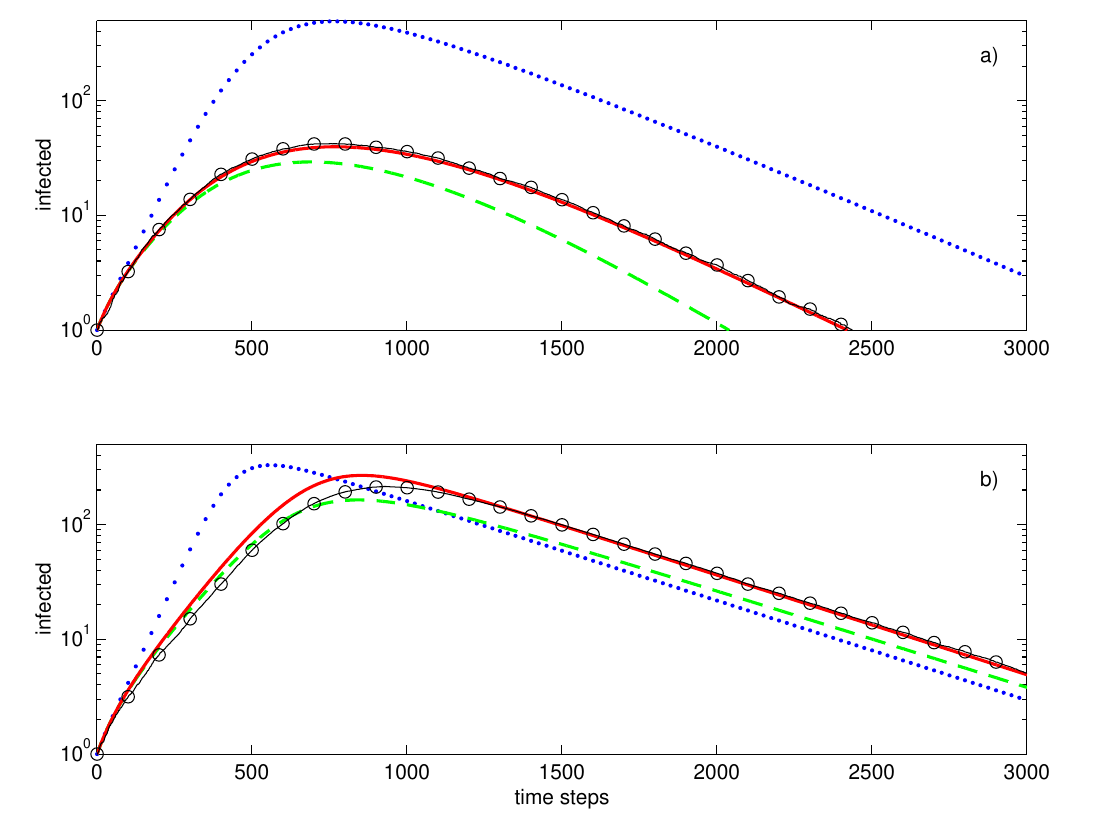}

\caption{Simulations of SIR epidemic on a tree (a), randomly generated graph (b) with a fixed degree of $d_i=4$ for every node   
Lines:      
$\bullet ~\bullet ~\bullet ~\bullet $ the first-order closure;
-- -- -- --    the second-order direct decoupling closure;
---------      the second-order conditional closure;
o---o---o      Monte-Carlo, ensemble averaging over 100 realisations. 
Simulation parameters: $\tilde{p}=0.005$,  $\tilde{q}=0.003$.  
\label{fig1-1} }

\end{center}
\end{figure}

The comparison of the closures with Monte-Carlo simulations is shown on Figure
\ref{fig1-1}a. The tree has 1457 nodes in 7 layers $l=0,...,6$. With exception
of the last (seventh) layer, each node has the degree of 4. Epidemic begins at
node $i=1$\ located at $l=0$. As expected the conditional closure, which is
exact in this case, is close to the average of the Monte-Carlo simulations.
Due to the need of evaluating multiple (100 in this case) realisations, the
Monte-Carlo simulations require a substantially longer computational time
(more than 30 times that of the closures). The second-order decoupling closure
has a noticeable error, while the first-order closure is substantially less
accurate than the closures of the second order.

\section{Modelling epidemic on scale-free networks}

The networks used in this section are created with assistance of random
generators, but the solutions are examined here for a fixed typical
realisation of each network, i.e. they are not averaged over possible
realisations of the networks. As in the previous sections, ensemble averaging
implies averaging over realisations of the stochastic simulations of the SIR
epidemic on a fixed network. The networks used here are scale-free and possess
small-world properties. These networks tend to increase the number of
accessible nodes exponentially with each infection transition to neighbours
--- this matches the initial exponential growth observed in most epidemics.
These exponents are strongly affected by the degrees of the nodes involved.

\begin{figure}
\begin{center}
\includegraphics[width=.75\linewidth,page=1,trim=1cm 0.5cm 2cm 20.5cm, clip ]{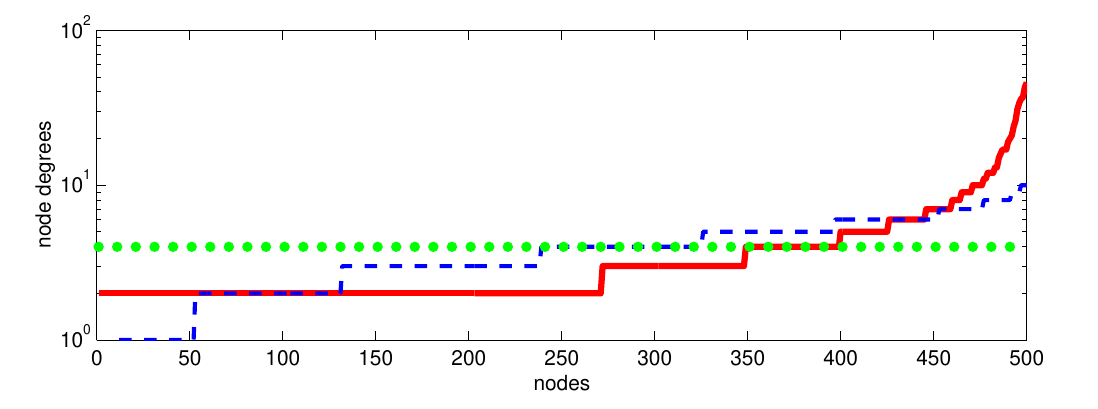}

\caption{Node degrees versus nodes (ordered by their degrees) for Erdős-Rényi (-- -- --), Barabasi-Albert (------) and random with a fixed degree 
($\bullet ~\bullet ~\bullet ~\bullet $) graphs used in simulations. \label{fig2} }

\end{center}
\end{figure}

All networks considered in this section have 500 nodes with the average degree
of 4 and, as shown in Figure \ref{fig2}, with rather different distributions
of node degrees. Figure \ref{fig1-1}a\ shows propagation of the SIR epidemic
on graph with connections between nodes selected at random constrained by the
requirement that degree of each node is exactly 4. At the initial stage, this
propagation is the same as propagation on a tree graph shown in Figure
\ref{fig1-1}a but as the number of infected nodes increases, the evolutions of
these epidemics diverge.

For the Erd\H{o}s--R\'{e}nyi graph --- the most simple random graph to
generate by connecting each couple of nodes with a given probability --- the
node degrees exhibit some random variations, which have the binomial
distribution. Another network, which is represented by Barab\'{a}si--Albert
scale-free graph and, as shown in Figure \ref{fig2}, has the largest
variations of the degrees, is considered to give a better representation of
the real-world networks. This graph is generated by adding new nodes
sequentially with random but preferentially distributed connections
proportional to the degrees of the existing nodes. This results in heavy
distribution tails: relatively few nodes have many connections. We call these
well-connected nodes "central" and the nodes with relatively few connections
"peripheral". While it can be argued that, if compared to real-world networks,
the Barab\'{a}si--Albert graphs tend to overestimate the heaviness of the
central nodes, this may be useful since the graphs used in simulations (which
have only 150-1500 nodes in the present work) are much smaller than millions
of susceptible agents in the real world, and exaggerated clustering\ of the
node degrees in small graphs realistically reflect the concentration of
connections in the real-world conditions.

The simulations are performed for the following values of the parameters
$\tilde{p}=p\Delta t=0.005$ and $\tilde{q}=q\Delta t=0.003,$ which are assumed
to be the same for all nodes. These values are sufficiently small to ensure
that simultaneous infection+recovery transitions are unlikely within the same
time step. The value of the time step is checked by reducing $\Delta t$ twice
and as expected, this does not affect the results. The time step should be
sufficiently small but not too small, as this increases computational
expenses. The transmission $p$ and recovery $q$ probabilities are selected to
provide a reasonable value for the $q/p$ ratio ensuring that transmission and
recovery have comparable magnitudes for the graphs examined here. The
Monte--Carlo implementation of the model conventionally generates pseudorandom
numbers determining stochastic transitions and, ultimately, the realisations
of the process.

\begin{figure}
\begin{center}
\includegraphics[width=.9\linewidth,page=1,trim=1cm 0.5cm 2cm 13.5cm, clip ]{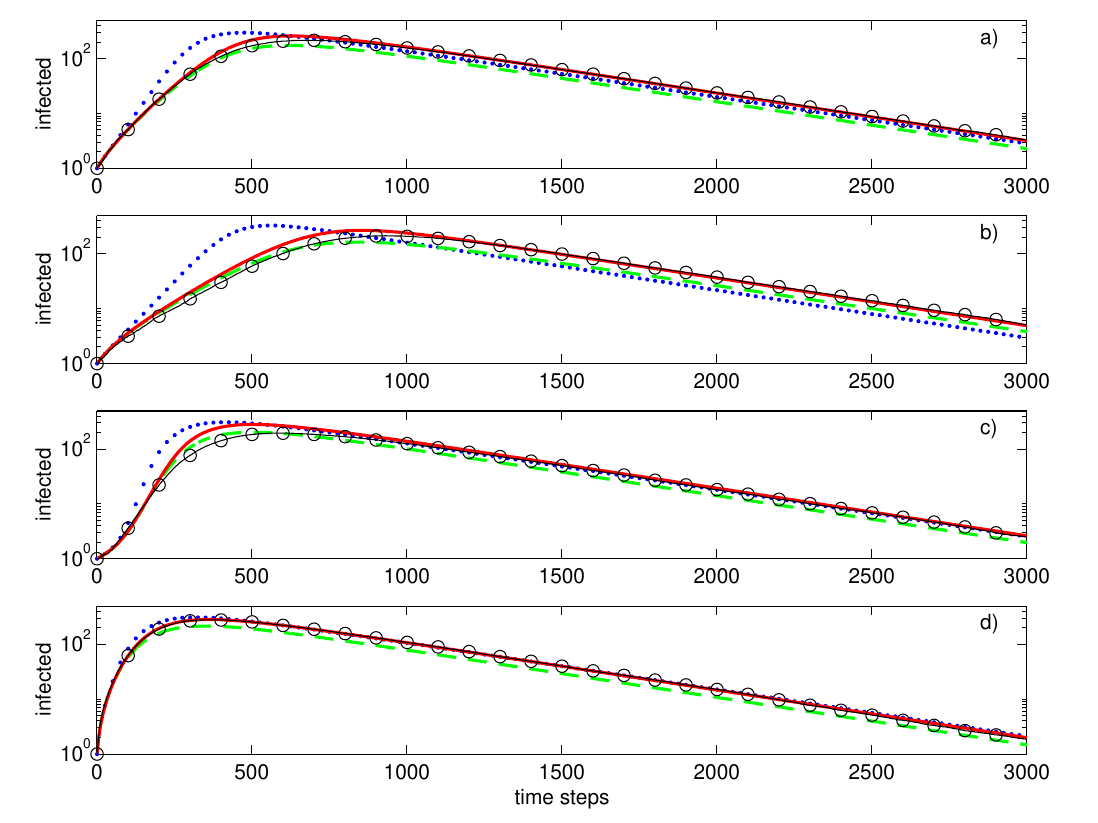}

\caption{Simulations of SIR epidemic on the Erdős-Rényi (a), fixed node degrees (b) and Barabasi-Albert (c,d) graphs with 
peripheral (c) and central (d) initial conditions.   
Lines:      
$\bullet ~\bullet ~\bullet ~\bullet $ the first-order closure;
-- -- -- --    the second-order direct decoupling closure;
---------      the second-order conditional closure;
o---o---o      Monte-Carlo, ensemble averaging over 100 realisations. 
Simulation parameters: $\tilde{p}=0.005$,  $\tilde{q}=0.003$.  
  \label{fig3} }

\end{center}
\end{figure}

Figure \ref{fig3}\ illustrates the outcomes of the simulations. The
first-order closure is less accurate than the second-order closures but is
still qualitatively correct. The random nature of the \ graphs tends to
increase chaos and decrease correlations between the nodes. Among the
second-order closures, the conditional closure is slightly better than the
direct decomposition and matches well the averages of stochastic simulations.
This averaging is evaluated over 100 independent realisations making
Monte-Carlo simulations relatively expensive.

The evolutions of the epidemic are substantially different for different
networks, even if all of these networks have the same average degree of 4. The
fixed degree network has the slowest development of the epidemic and the most
stable value of the growth exponent. The epidemic progresses faster for the
Barab\'{a}si--Albert network. The Erd\H{o}s--R\'{e}nyi network demonstrates
behaviour that is intermediate between that of Barab\'{a}si--Albert and fixed
degree networks. Note that the growth exponent is not constant for networks
that have significant variations of the node degrees. This is most evident for
the Barab\'{a}si--Albert network, which demonstrates the largest slope of the
exponent followed by its subsequent reduction. This network has another effect
associated with the initial conditions: whenever the initial node igniting the
epidemic is peripheral, there is a substantial delay in the evolution of the
epidemic (as illustrated by Figure \ref{fig3}c in comparison with Figure
\ref{fig3}d).

\begin{figure}
 \captionsetup[subfigure]{labelformat=empty}
\begin{center}

\subfloat{\frame{\includegraphics[width=.32\linewidth,page=1,trim=4.1cm 2cm 4.2cm 14cm, clip  ]{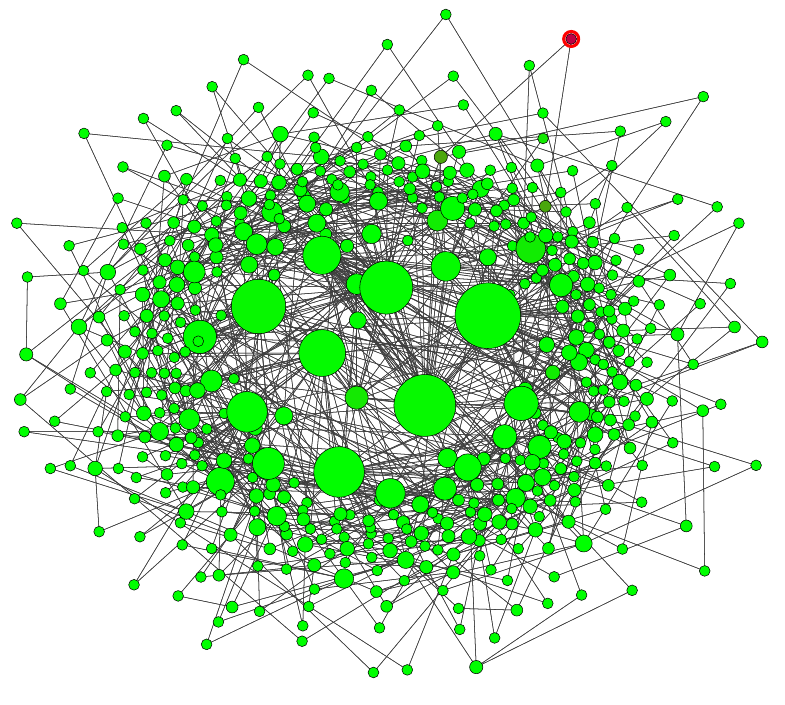}}}
\subfloat{\frame{\includegraphics[width=.32\linewidth,page=1,trim=4.1cm 2cm 4.2cm 14cm, clip  ]{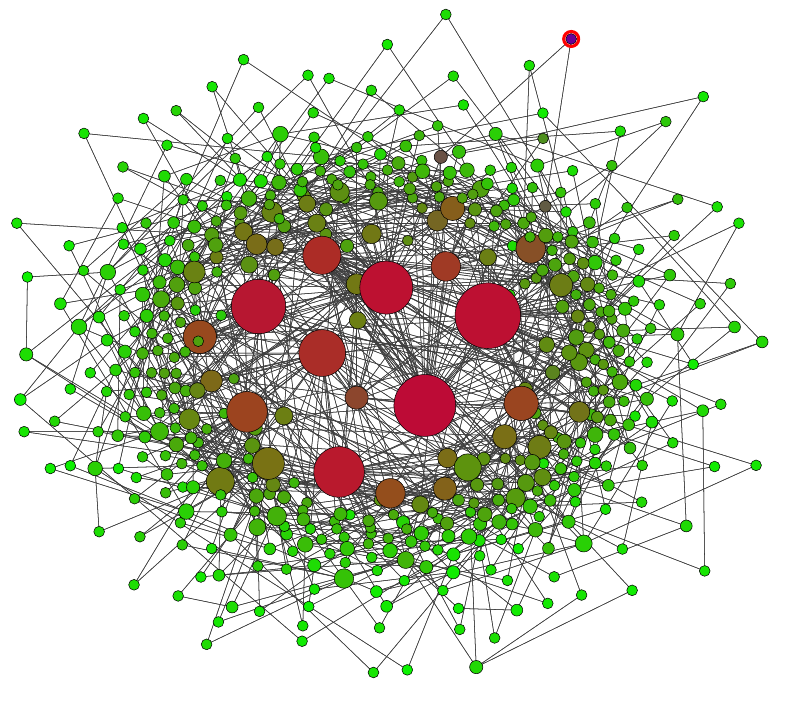}}}
\subfloat{\frame{\includegraphics[width=.32\linewidth,page=1,trim=4.1cm 2cm 4.2cm 14cm, clip  ]{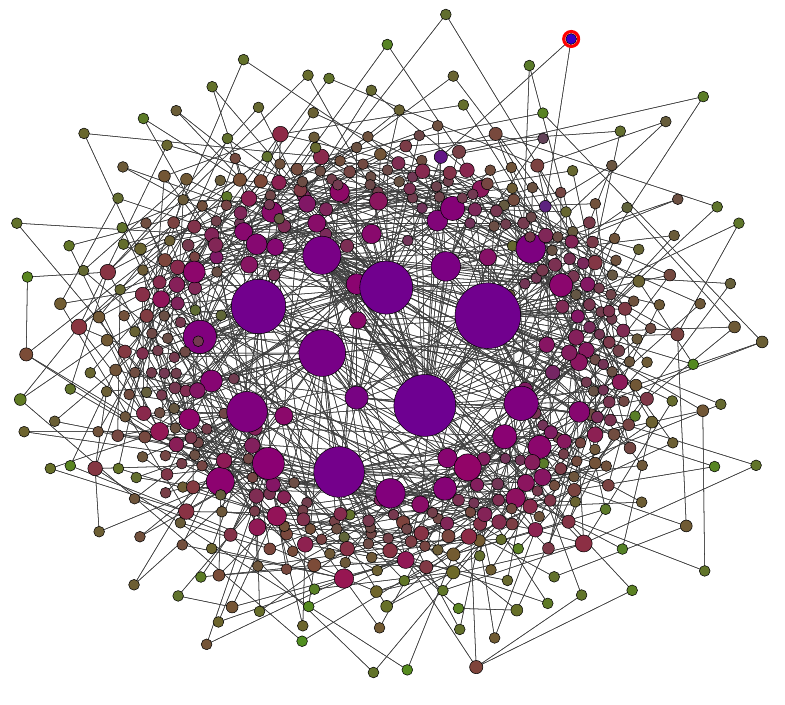}}}

\subfloat{\frame{\includegraphics[width=.32\linewidth,page=1,trim=4.1cm 2cm 4.2cm 14cm, clip  ]{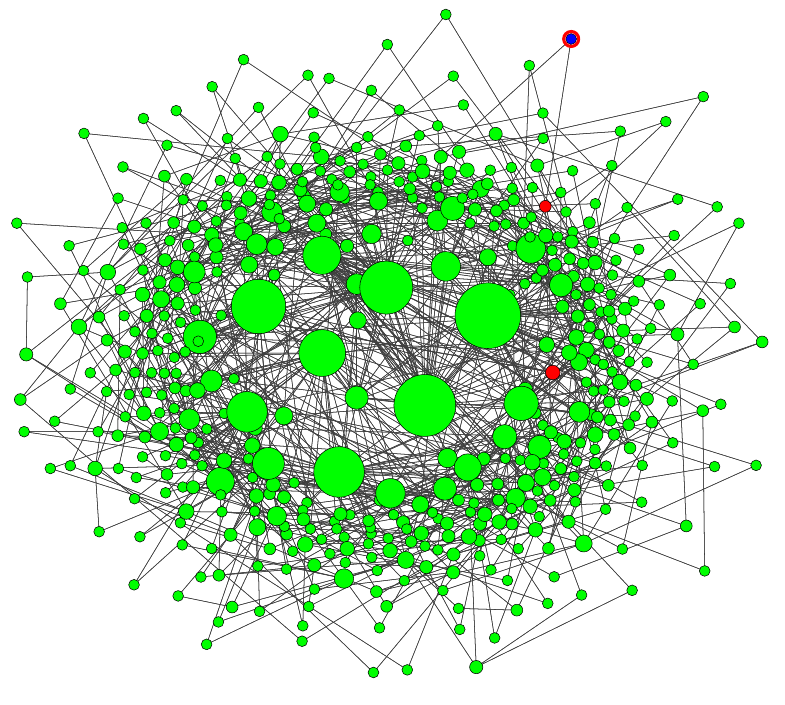}}}
\subfloat{\frame{\includegraphics[width=.32\linewidth,page=1,trim=4.1cm 2cm 4.2cm 14cm, clip  ]{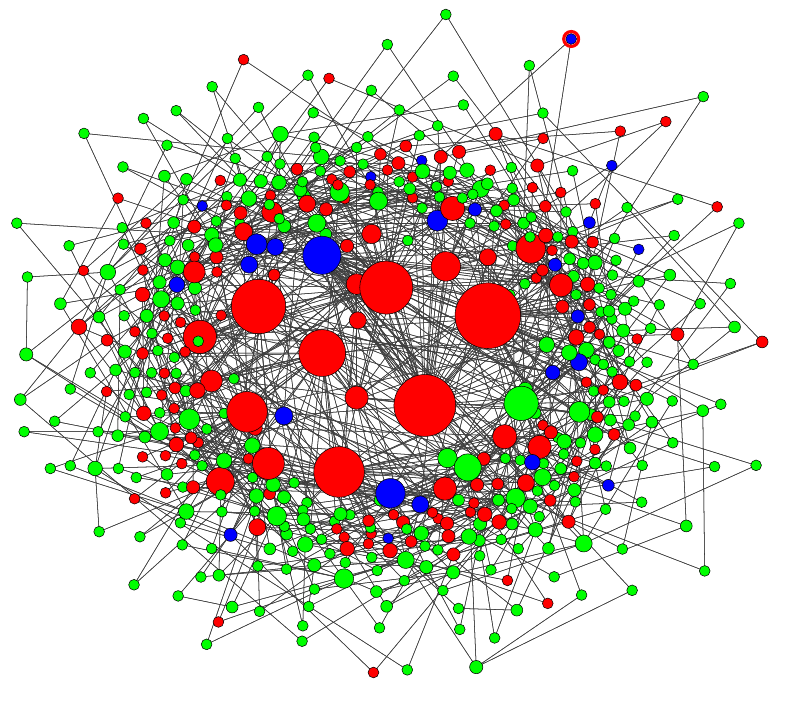}}}
\subfloat{\frame{\includegraphics[width=.32\linewidth,page=1,trim=4.1cm 2cm 4.2cm 14cm, clip  ]{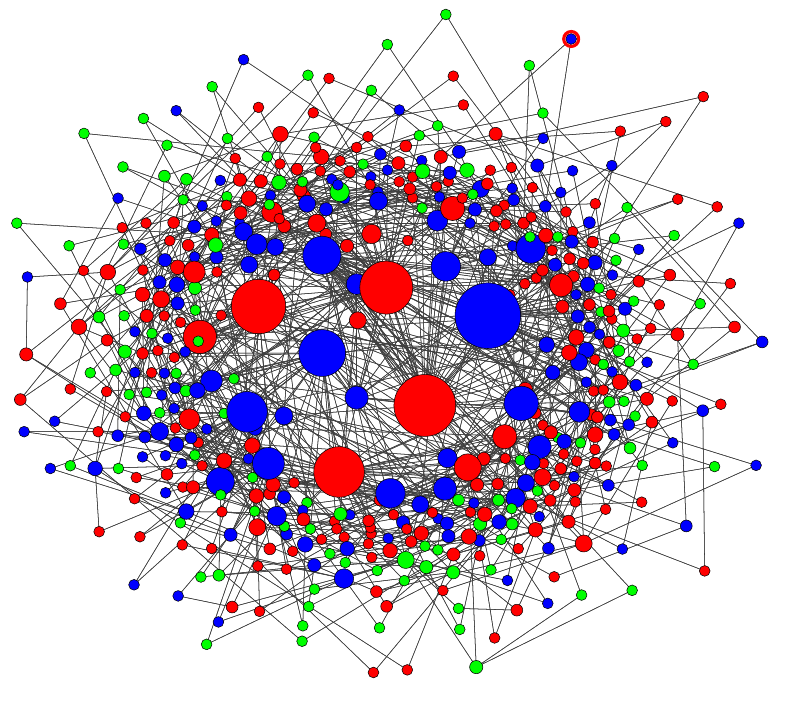}}}

\subfloat{\frame{\includegraphics[width=.32\linewidth,page=1,trim=4.1cm 2cm 4.2cm 14cm, clip  ]{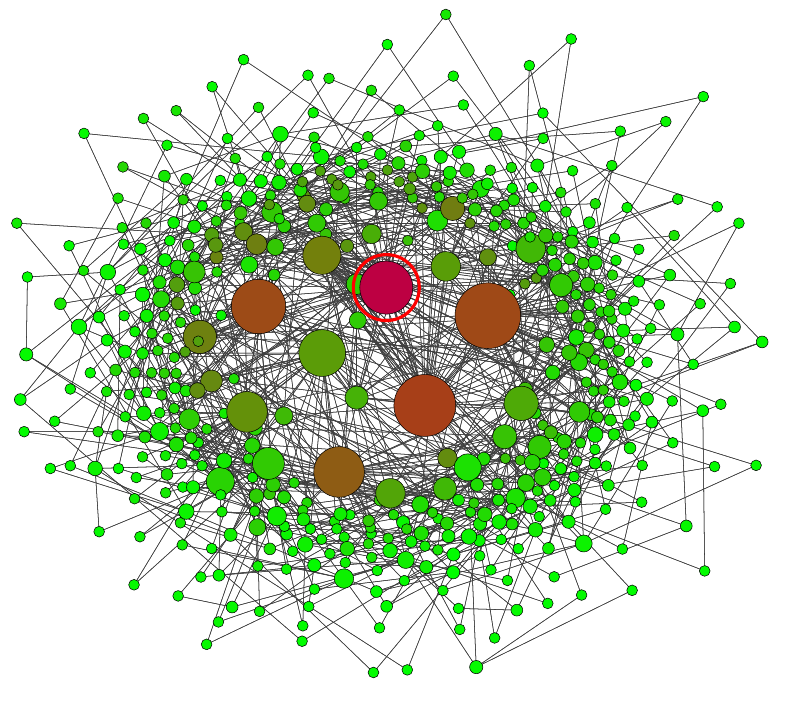}}}
\subfloat{\frame{\includegraphics[width=.32\linewidth,page=1,trim=4.1cm 2cm 4.2cm 14cm, clip  ]{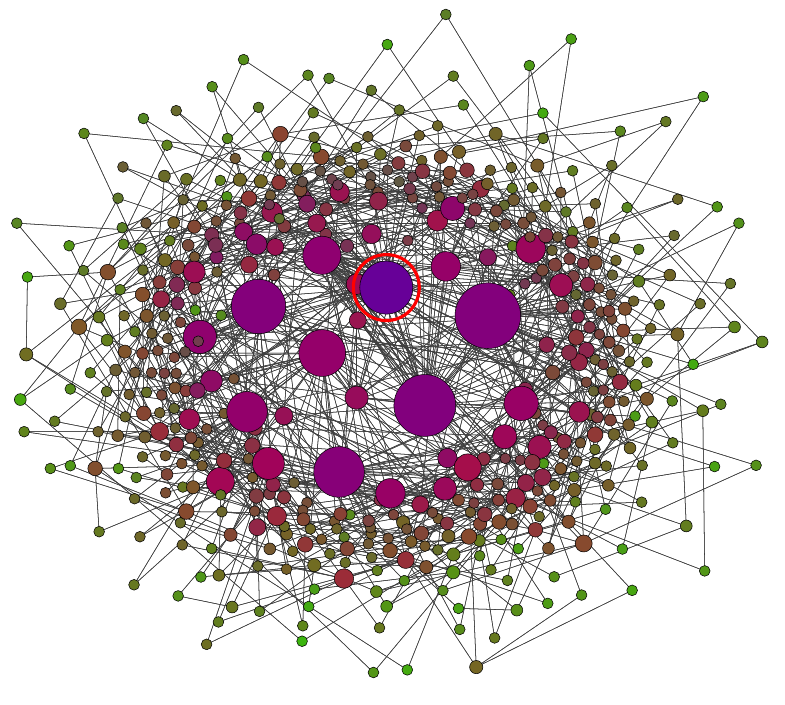}}}
\subfloat{\frame{\includegraphics[width=.32\linewidth,page=1,trim=4.1cm 2cm 4.2cm 14cm, clip  ]{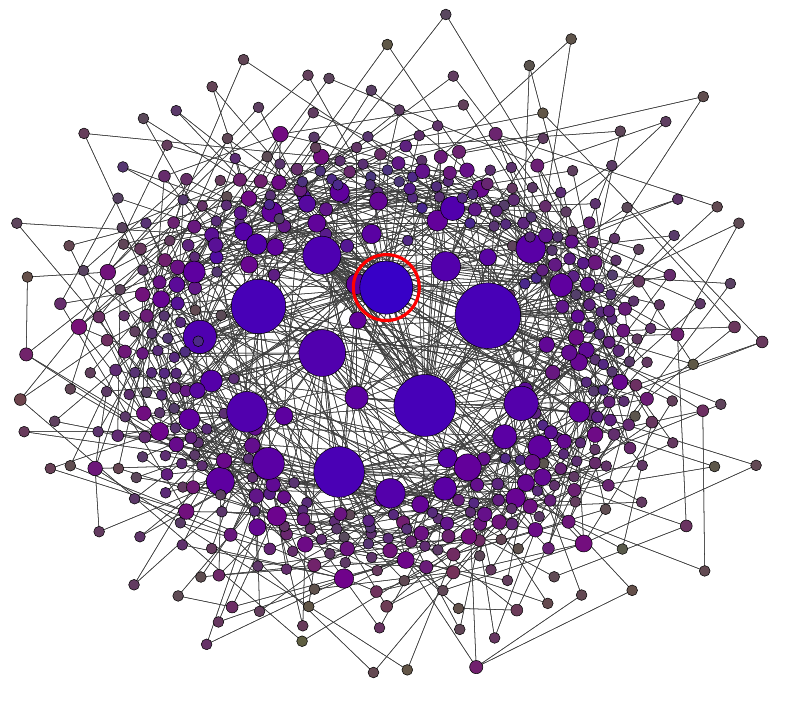}}}

\subfloat[100 time steps]{\frame{\includegraphics[width=.32\linewidth,page=1,trim=4.1cm 2cm 4.2cm 14cm, clip  ]{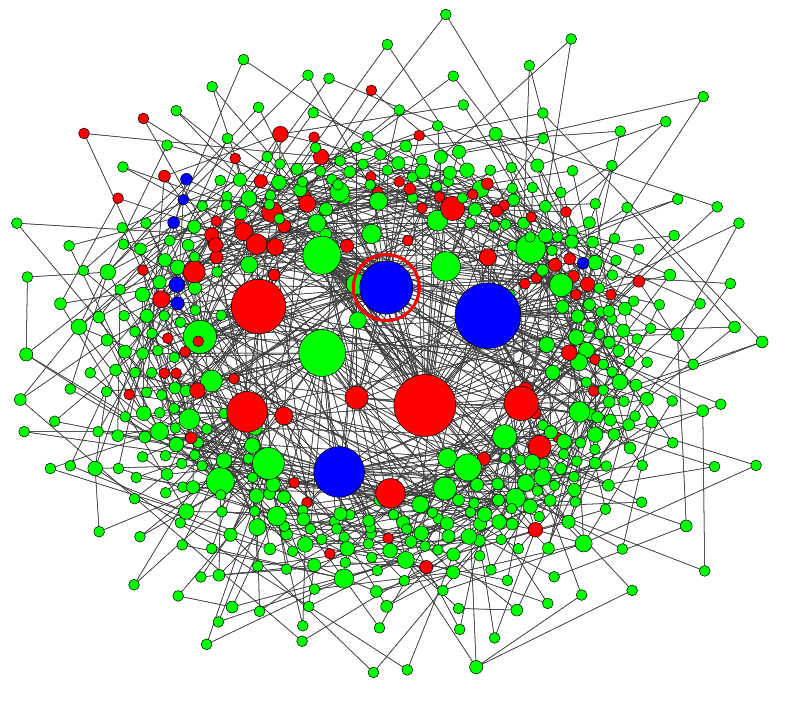}}}
\subfloat[300 time steps]{\frame{\includegraphics[width=.32\linewidth,page=1,trim=4.1cm 2cm 4.2cm 14cm, clip  ]{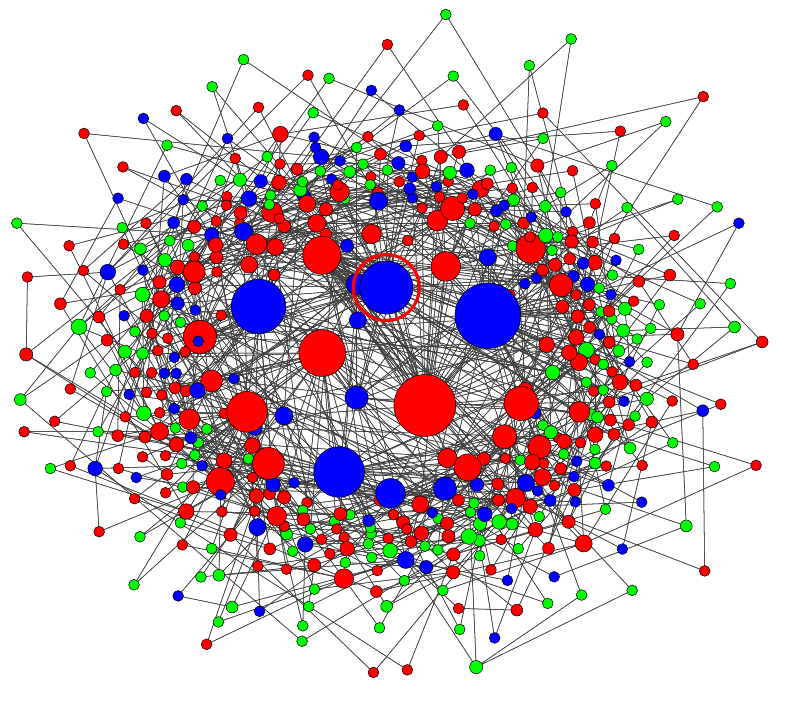}}}
\subfloat[500 time steps]{\frame{\includegraphics[width=.32\linewidth,page=1,trim=4.1cm 2cm 4.2cm 14cm, clip  ]{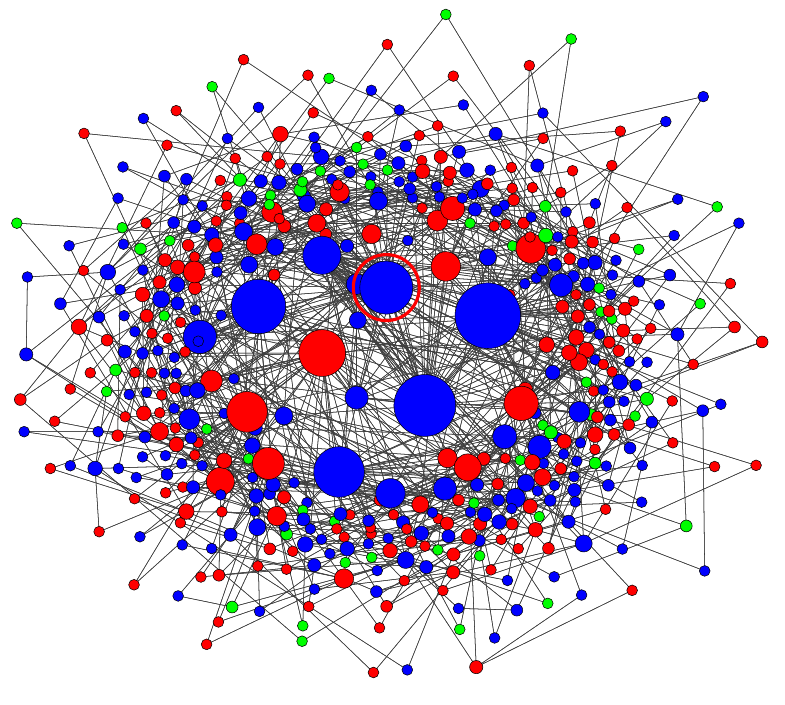}}}


\caption{SIR model on the Barabási–Albert network of 500 nodes. Columns from left to right: 100, 300  and 500 time steps. 
Rows (from top to bottom): 1,2 -- peripheral initial condition; 3,4 -- central initial condition; 1,3 --  conditional closure model; 2,4 -- Monte-Carlo simulations. The initially infected nodes are indicated by the red circles. The node sizes are proportional to the node degrees. Node colours: green -- susceptible, red -- infected, blue -- recovered. Simulation parameters: $\tilde{p}=0.005$,  $\tilde{q}=0.003$.   
 \label{fig4}}

\end{center}

\end{figure}

Figure \ref{fig4} shows the Barab\'{a}si--Albert network at different stages
of the epidemic. Infection of a central node (Figure \ref{fig4}c and d) leads
to fast propagation of the infection to the whole central segment followed by
a slower expansion of the epidemic to the peripheral nodes. Initial infection
of a peripheral node (Figure \ref{fig4}a and b) results in a substantial delay
in infecting the central segment. During this delay, the epidemic remains
latent and, occasionally, might become extinct. Once infection reaches a
central node, the rate of infection accelerates dramatically. Note that
occasional (local) extinctions amplify correlations between neighbouring nodes
and, as this is well-known in modelling of reacting flows, increase the
complexity of the simulations.

\bigskip\pagebreak



\section{Conclusion}

This study evaluates the application of a statistical mechanics-based
framework, utilising agent-based Susceptible-Infected-Recovered (SIR) models
formulated as continuous-time Markov processes on networks, with the primary
aim of testing the effectiveness of combining this approach with complex
network geometry. The methodology involves deriving a hierarchical system of
marginal probability equations, analogous to the BBGKY hierarchy, to capture
complex stochastic transitions and network-driven dependencies within disease spread 
(although illustrative results are often reported in terms of aggregate averages).

Findings indicate that the second-order conditional closure yields a closed
system of equations that approximates Monte Carlo simulations with reasonably
high fidelity, particularly in complex topologies, associated with network
clustering. The approach can effectively reproduce some notable features of
Monte-Carlo-simulated epidemic propagation and extinction, as well as the
influence of network structure and possible intervention measures, such as
lockdowns. The model also appears to reproduce some qualitative aspects of the
COVID-19 epidemic in Lombardy: it specifically identified the persistence of
infection despite high average herd immunity pointing to the roles of
community clustering. However, this application is intended primarily as a
methodological assessment---testing the analytical and computational
properties of the proposed closure technique---rather than as a comprehensive
representation of real-world epidemic processes.

The results further suggest that this statistical mechanics and Markov process
approach has potential utility in modelling diffusion phenomena beyond
epidemiology, including the adoption of new energy technologies and responses
to climate-related disruptions. Future research should address expanded
health-state architectures, adaptive network features, and heterogeneous agent
behaviours to enhance the model's capacity for representing multi-phase
processes and behavioural feedbacks.

Overall, the study provides evidence supporting the efficacy of hierarchical
closure techniques for network-based epidemic modelling, especially at the
point when conventional "chaotic" assumptions break and complexity emerges.
Its application demonstrates that the framework is applicable to a broad class
of diffusion and cascade processes in the presence of complex structures and
interlinks. This work establishes a baseline for future methodological
refinement and for cross-domain applications of hierarchical modelling of
competitive diffusion and emergent complexity in the presence of network
heterogeneity and clustering.

\appendix

\section*{Appendix A: Network clustering and epidemic waves}

\label{app:clustering}

Real-world contact networks are rarely homogeneous: they typically exhibit
community structure (clusters) and, often, a hierarchy of subclusters. Such
clustering can strongly modify epidemic dynamics and may generate multi-wave
behaviour even in simple agent-based SIR settings on static graphs; see, e.g.,
network-based studies in computational epidemiology
\cite{DellaRossa2020,Pizzuti2020,Lombardi2021}. In particular, once the
infection has largely saturated the central (high-degree) nodes of one
community and local herd immunity begins to form, the epidemic may
nevertheless persist if the infection subsequently reaches the central nodes
of other communities. This mechanism provides a natural route to prolonged
tails and secondary waves \cite{kermack1927contribution}. 
Note that communities is a mathematical term describing clustering in networks. Such communities may coincide with the everyday meaning of the term (here we refer to overt communities, such as those associated with geographical location), but they may also be latent—not physically separated and not directly observable. Consequently, transmission between communities may or may not be evident in practice, and an intrinsically inhomogeneous process can appear homogeneous in aggregate data. 

These subtle divisions between communities may become more pronounced in modern societies as people are better informed and adjust their behaviour, while governments introduce emergency regulations. 
Measures commonly described as \textquotedblleft lockdowns\textquotedblright\ can be
represented, at a minimal level, by a reduction in the effective propagation
probability $p$ and/or by reduced inter-community mixing; in clustered
networks this can temporarily delay spill-over between communities but does
not, by itself, eliminate the possibility of later re-amplification .

\begin{figure}[t]
\begin{center}
\includegraphics[width=.75\linewidth,page=1, trim=1cm 7.5cm 2cm 13.5cm, clip ]{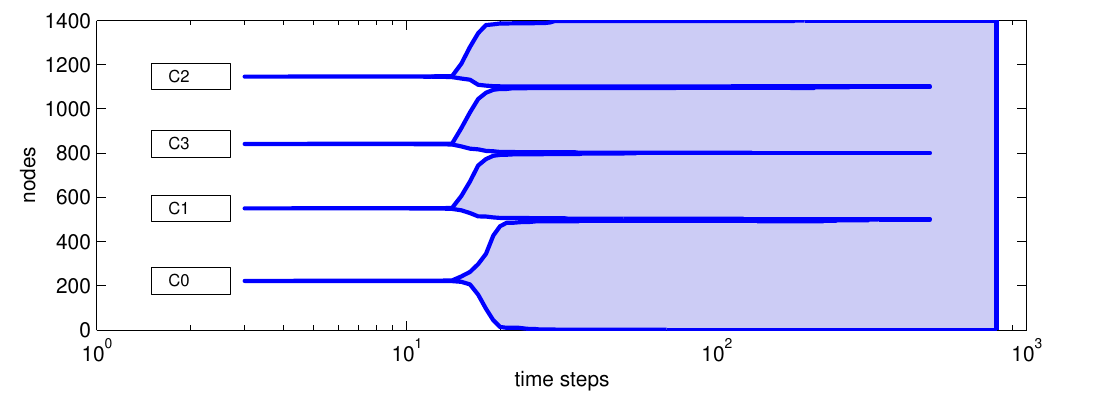}
\caption{Clustering dendrogram for the composite network \label{figA4}}
\end{center}
\end{figure}

\begin{figure}[h]
 \captionsetup[subfigure]{labelformat=empty}
\begin{center}

\subfloat{\frame{\includegraphics[width=.45\linewidth,page=1,trim=3.3cm 9cm 4cm 10cm, clip  ]{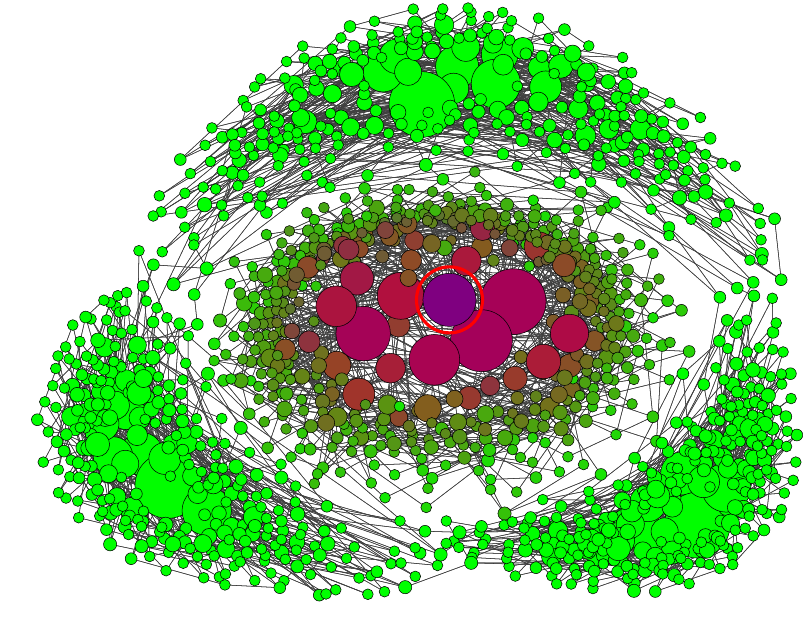}}}
\subfloat{\frame{\includegraphics[width=.45\linewidth,page=1,trim=3.3cm 9cm 4cm 10cm,clip  ]{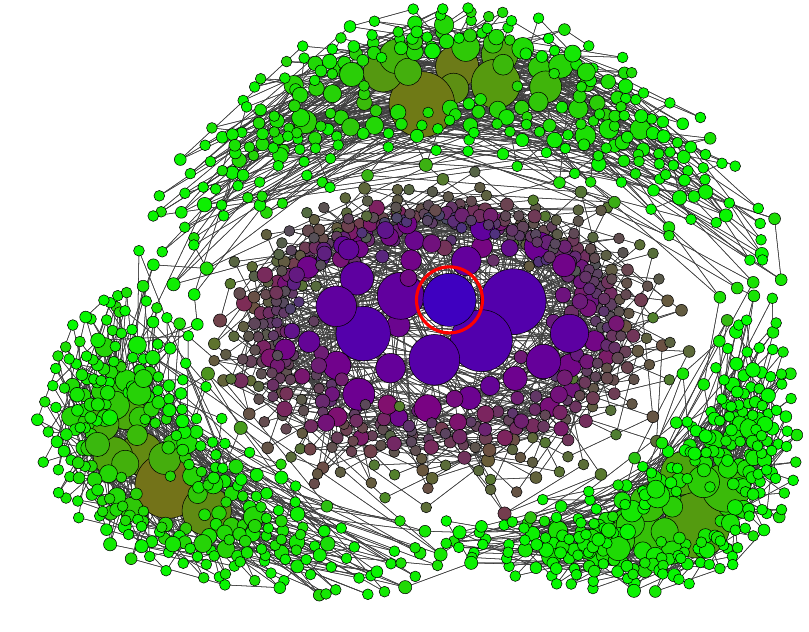}}}

\subfloat[200 time steps]{\frame{\includegraphics[width=.45\linewidth,page=1,trim=3.3cm 9cm 4cm 10cm, clip  ]{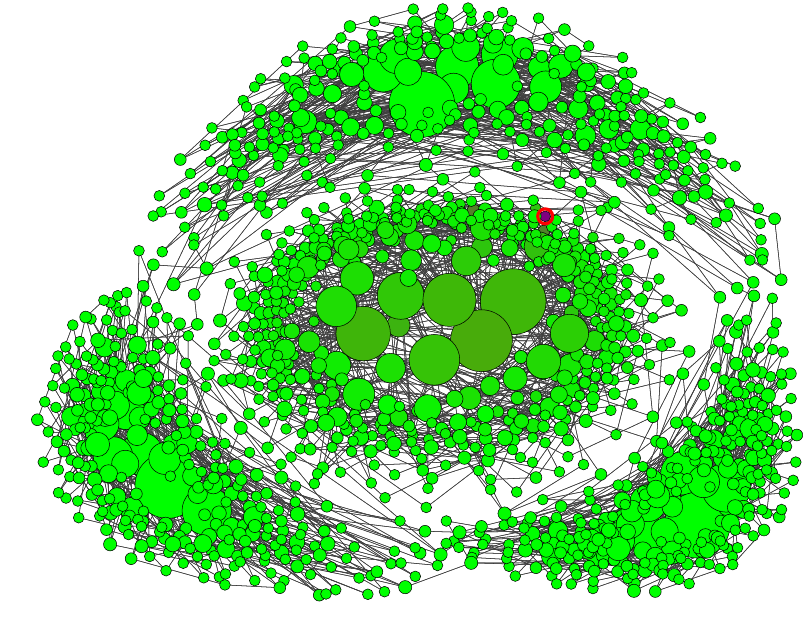}}}
\subfloat[400 time steps]{\frame{\includegraphics[width=.45\linewidth,page=1,trim=3.3cm 9cm 4cm 10cm, clip  ]{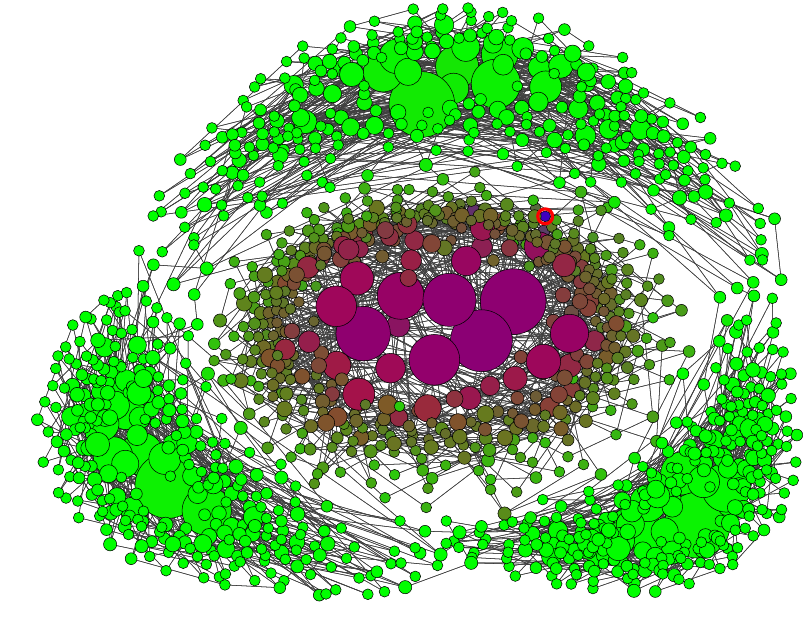}}} \\


\caption{SIR model with the conditional closure on the composite network of 4 communities at 200 time steps (left) 
 and 400 time steps (right). 
The initially infected node is indicated by a red circle and is positioned centrally (top row) or peripherally (bottom row). The node sizes are proportional to the node degrees.  Node colours: green -- Susceptible, red -- Infected, blue -- Recovered. Mixed colours indicate probabilities of S,I and R. Simulation parameters: $\tilde{p}=0.005$,  $\tilde{q}=0.003$.  
 \label{figA3}}

\end{center}

\end{figure}

\begin{figure}[h]
\begin{center}
\includegraphics[width=.75\linewidth,page=1,trim=1cm 0.5cm 2cm 20.5cm, clip ]{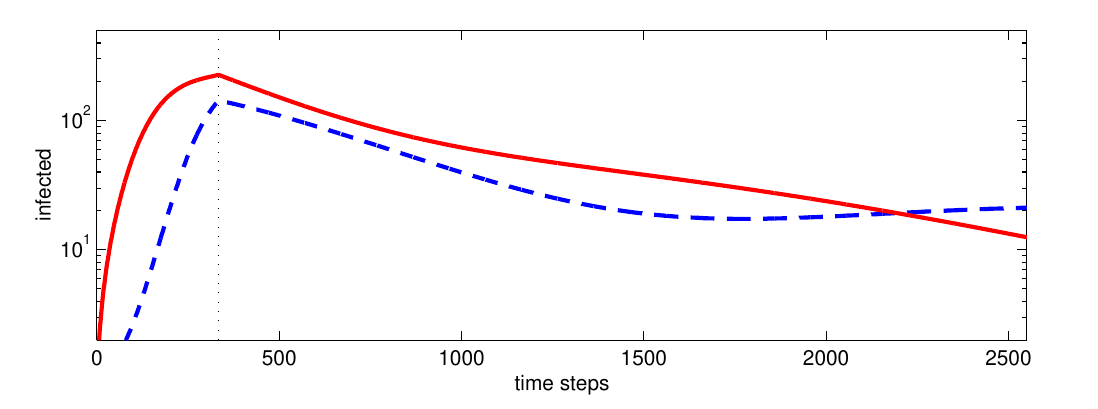}

\caption{Number of infected nodes versus time step for epidemic simulation using conditional closure and with a lockdown at 333 time steps indicated by the vertical line. Simulation parameters: $\tilde{p}=0.005$  $\tilde{q}=0.0035$ while $\tilde{p}=0.001$ after lockdown.   Initial condition: a) central (------)  , b) peripheral (-- -- --).     \label{figA5}}

\end{center}
\end{figure}

To illustrate this effect, we consider a composite network formed by four
Barab\'asi--Albert graphs: one principal community (500 nodes) and three
secondary communities (300 nodes each), weakly connected so that their central
segments do not merge. The clustering dendrogram in Figure~\ref{figA4}
confirms the presence of four communities \cite{IMA2019}. Figure~\ref{figA3}
shows a representative evolution in which the epidemic first infects the
principal community (``metropolis''), begins to subside there as immunity
accumulates, yet escapes into secondary communities and produces renewed
growth. Figure~\ref{figA5} compares central and peripheral initial conditions
on the same clustered network and demonstrates that the post-intervention
evolution can be governed by competing trends: attenuation within the
initially affected community versus delayed ignition of other communities,
which can generate secondary waves or long-lasting plateaux.

\begin{figure}[h]
\begin{center}
\includegraphics[width=.75\linewidth,page=1,trim=1cm 0.5cm 2cm 20.5cm, clip ]{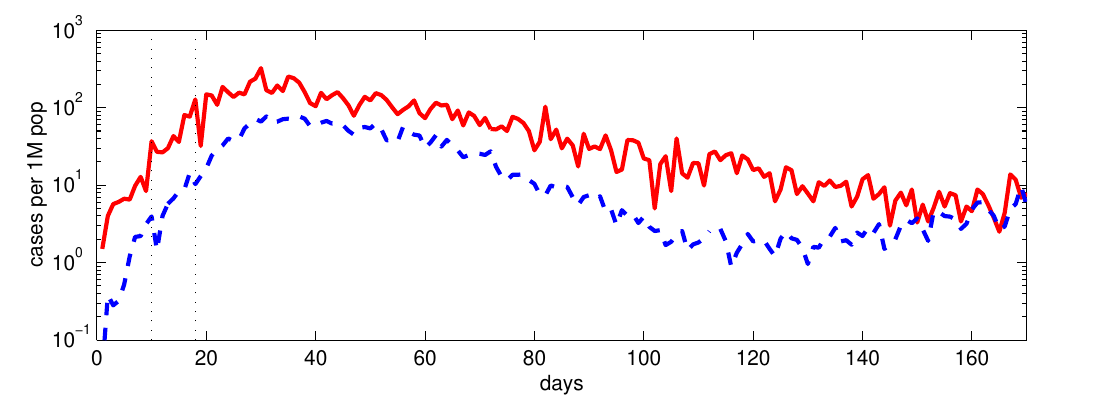}

\caption{Covid infections per 1 million population in Italy during the first covid wave, February-June, 2020.  
Lines a) ------ Lombardy,  b) -- -- -- the rest of Italy (excluding Lombardy) 
The vertical lines indicate the lockdown time and the effective delay associated with the latent period.    
\label{figA2}}

\end{center}
\end{figure}

A qualitatively similar pattern may have been present during the first wave of the COVID-19 epidemic in Italy. As illustrated in Figure~\ref{figA2}, the decline in Lombardy appears slower than in neighbouring regions despite a larger initial burden \cite{AnnaOdone2020CdiL} and, therefore, presumably higher immunity, at least in some segments of the population. This observation is not straightforward to reconcile with a single well-mixed community, where deeper penetration of the infection would typically be expected to accelerate the subsequent decay.

While many factors may contribute, clustered-network dynamics provide a natural explanation. Strong early spread within one community—potentially triggered by an unfortunate early infection of highly connected (central) individuals—can coexist with delayed propagation into other communities. Deep penetration then has two competing effects: it builds immunity within the initially affected community, while also seeding infection more widely across other communities. As incidence declines in the first community owing to local herd immunity, transmission initiated elsewhere can sustain the overall level of infection, producing persistence and a prolonged tail without necessarily developing into a distinct second wave. The qualitative similarity between the trajectories in Figures~\ref{figA5} and \ref{figA2} supports the plausibility of this interpretation.

\section*{Appendix B: Nomenclature}


\noindent\hspace*{6.5cm}\begin{minipage}{\dimexpr\textwidth-6.5cm\relax}
\begin{description}
\setlength{\labelwidth}{5.5cm}   
\setlength{\leftmargin}{\labelwidth}
\addtolength{\leftmargin}{0.5cm}
\setlength{\labelsep}{0.5cm}
\item[$Y_i\in\{S,I,R\}$] SIR states
\item[$Y^{(N)}=Y_{1},\ldots,Y_{N}$] System state vector
\item[$P_Y$] Full joint probability
\item[$\langle\cdot\rangle$] Ensemble average
\item[$\theta_i(Y_i^{\circ})=\delta_{Y_i Y_i^{\circ}}\in\{0,1\}$] Indicator function
\item[$f^{(n)}=\theta_{i_1}(\cdot)\cdots\theta_{i_n}(\cdot)$] Fine-grained distribution
\item[$P^{(n)}=\langle f^{(n)}\rangle=\langle\theta_{i_1}\cdots\theta_{i_n}\rangle$] Marginal probability
\item[$A_{ij}$, \textnormal{with} $A_{ij}=A_{ji}$ \textnormal{and} $A_{ii}=0$] Adjacency matrix
\item[$T$] Transition-rate operator
\item[$S\rightarrow I\rightarrow R$] Infection / recovery transitions
\item[$p_i$] Infection parameter
\item[$q_i$] Recovery parameter
\item[$\Phi_i=\sum_{j} p_i A_{ij}\,\theta_i(S)\,\theta_j(I)$] Infection transition rate (operator)
\item[$\Psi_i=q_i\,\theta_i(I)$] Recovery transition rate (operator)
\item[$\bar{\Phi}_i,\ \bar{\Psi}_i$] Averaged transition rates
\end{description}
\end{minipage}

\bibliographystyle{unsrt}
\bibliography{SIRs}

\end{document}